\documentclass[12pt]{amsart}
\usepackage{epsfig}
\begin{document}
\topmargin=10mm
 \oddsidemargin=0mm
 \evensidemargin=0mm
\baselineskip=20pt
\parindent 20pt
\begin{center}
{\Large\bf  Constructing periodic  wave solutions of nonlinear
equations by Hirota bilinear method }\\[12pt]
   Huihui Dai$^{a}$,\ \ Engui Fan$^{b}$, \ \  Xianguo Geng$^{c}$\\[8pt]
 {\footnotesize a. Department of Mathematics, City University of Hong Kong,  Hong
Kong SAR, China \\[6pt]
 b.  Institute of Mathematics,  Fudan University, Shanghai 200433, P. R. China.
\\
 c. Department of Mathematics, Zhengzhou University, Zhengzhou, P. R. China}\\[12pt]
\end{center}
{\bf Abstract.} Hirota bilinear method is proposed  to directly
construct periodic wave solutions in terms of Riemann theta
functions for nonlinear equations.  The asymptotic property of
periodic waves are analyzed in detail. It is shown that well-known
soliton solutions can be reduced from the periodic wave solutions.
\\[12pt]
 {\bf 1. Introduction}

  The investigation of the exact solutions of
nonlinear equations plays an important role in the study of
nonlinear physical  phenomena. For example, the wave phenomena
observed in fluid dynamics, plasma and elastic media are often
modelled by  the bell shaped sech solutions and the kink shaped
tanh traveling wave solutions. The exact solution,  if available,
of those nonlinear equations facilitates the verification of
numerical solvers and aids in the stability analysis of solutions.
In the past decades,  there has been significant progression in
the development of these methods  such as inverse scattering
method [1,2],  Darboux transformation [3-6], Hirota bilinear
method [7-12], algebro-geometric method [13-18] and others. Among
them, the algebro-geometric method is an analogue of inverse
scattering transformation, which was first developed by Matveev,
Its, Novikov, and Dubrovin et al. The method can derive an
important class of exact solutions, which is called quasi-periodic
or algebro-geometric solution, to many soliton equations such as
KdV equation, sin-Gordon equation, and Schrodinger equation. In
recent years, such solutions of nonlinear equations have been
aroused much interest in the mathematical physics [19-24].
However, this method usually is applied in the integrable
nonlinear evolution equations admitting Lax pairs representation
and involves complicated algebraic geometry theory. These have
been used far less than their soliton counterparts. The main
reason for this is that they are far more complicated. Soliton
solutions are typically expressed in terms of rational or
hyperbolic functions, whereas qusi-periodic solutions require the
use of Riemann theta functions and calculus on Riemann surfaces.
Recently, Deconinck, Hoeij et al proposed an algorithm to compute
the Riemann theta function and Riemann constants [25,26].

  It is well known that the bilinear derivative method developed by Hirota is a
powerful and direct approach to construct exact solution of
nonlinear equations [7-9]. Once a nonlinear equation is written in
bilinear forms by a dependent variable transformation, then
multi-soliton   solutions and rational solutions can be obtained.
In recent years,  Hirota  method also has been developed for
obtaining Wronskian and Pfaffian forms of N-soliton solution
[10,11]. In this paper, we propose Hirota method for directly
constructing  periodic wave solutions in Riemann theta functions.
It is shown that the periodic solutions can be reduced to
classical soliton solutions under a certain limit. The appeal and
success of this method lies in the fact we circumvent complicated
algebro-geometric theory to directly get explicit periodic wave
solutions.  Moreover, all parameters appearing in the solutions
are free variables, whereas  usual quasi-periodic solutions
involve  Riemann constants which are difficult to be determined
 and need to make complicated Abel transformation on Riemann surface [13-18].
 As illustrative examples, we then consider KdV  equation and KP equation in this paper.
 Other many equations also can be dealt with this way.\\[12pt]
{\bf 2.  Periodic wave solutions of KdV equation and their
reduction}

 Consider KdV equation
$$u_t+6uu_x+u_{xxx}=0.\eqno(2.1)$$
Substituting transformation
$$u=u_0+2(\ln f)_{xx}\eqno(2.2)$$
into Eq.(2.1) and integrating once again, we then get the
following  bilinear form
$$G(D_x,D_t)f\cdot f=(D_xD_t+6u_0D_x^2+D_x^4+c)f\cdot f=0\eqno(2.3)$$
where $c=c(t)$ is a integration constant and $u_0$ is a constant
solution of KdV equation.

The Hirota bilinear differential operator is defined by
$$D_x^mD_t^n f(x,t)\cdot g(x,t)=(\partial_x-\partial_{x'})^m(\partial_t-\partial_{t'})^n
f(x,t) g(x,t)|_{x=x',t=t'}$$
  The D-operator have good property when acting on exponential
  functions
  $$D_x^mD_t^n e^{\xi_1}\cdot
  e^{\xi_2}=(k_1-k_2)^m(\omega_1-\omega_2)^n
  e^{\xi_1+\xi_2}$$
where $\xi_j=k_jx+\omega_jt, j=1,2$. More general, we have
$$G(D_x,D_t)e^{\xi_1}\cdot
  e^{\xi_2}=G(k_1-k_2,\omega_1-\omega_2)
  e^{\xi_1+\xi_2}\eqno(2.4)$$

We consider Riemann theta  function solution of KdV equation
$$f=\sum_{n\in Z^N}e^{\pi i<\tau n,n>+2\pi i<\xi,n>}\eqno(2.5)$$
where $n=(n_1,\cdots,n_N), \xi =(\xi_1, \cdots, \xi_N)$, $\tau$ is
a symmetric matrix and ${\rm Im}\tau>0$, $\xi_j=k_jx+\omega_j t,
j=1,\cdots, N.$

Consider the case when $N=1$, then (2.5) becomes
$$f=\sum_{n=-\infty}^{\infty}e^{2\pi in\xi+\pi in^2\tau}\eqno(2.6)$$

Substituting (2.6) into (2.3) gives
\begin{eqnarray*}
&&Gf\cdot f=G(D_x,D_t)\sum_{n=-\infty}^{\infty}e^{2\pi in\xi+\pi
in^2\tau}\cdot\sum_{m=-\infty}^{\infty}
e^{2\pi i m\xi+\pi i m^2\tau}\\
&&=\sum_{n=-\infty}^{\infty}\sum_{m=-\infty}^{\infty}G(D_x,D_t)e^{2\pi
in\xi+\pi in^2\tau}\cdot e^{2\pi i m\xi+\pi i
m^2\tau}\\
&&=\sum_{n=-\infty}^{\infty}\sum_{m=-\infty}^{\infty}G(2\pi
i(n-m)k,2\pi i(n-m)\omega)e^{2\pi i(n+m)\xi+\pi i ( n^2+m^2)\tau}\\
&&\stackrel{n+m=m'}{=}\sum_{m'=-\infty}^{\infty}\left\{
\sum_{n=-\infty}^{\infty}G(2\pi i(2n-m')k,2\pi
i(2n-m')\omega)e^{\pi
i[(n^2+(n-m')^2]\tau}\right\}e^{2\pi i m'\xi}\\
&&\equiv\sum_{m'=-\infty}^{\infty}\bar{G}(m')e^{2\pi i m'\xi}=0.
\end{eqnarray*}

Noticing that
\begin{eqnarray*}
&&\bar{G}(m')=\sum_{n=-\infty}^{\infty}G(2\pi i(2n-m')k,2\pi
i(2n-m')\omega)e^{\pi i[(n^2+(n-m')^2]\tau}\\
&&\stackrel{n=n'+1}{=}\sum_{n'=-\infty}^{\infty}G(2\pi
i[2n'-(m'-2)]k,2\pi i[2n'-(m'-2)]\omega)\\
&&\ \ \  \  \ \times \exp(\pi i[(n'^2+(n'-(m'-2))^2]\tau)
\exp(2\pi i(m'-1)\tau)\\
&&=\bar{G}(m'-2)e^{2\pi i (m'-1)\tau}\\
&&=\cdots=\left\{\begin{matrix}\bar{G}(0)e^{\pi im'(m'-1)},&m' \ \
{\rm is\ \ even}\cr\cr \bar{G}(1)e^{\pi i(m'+1)(m'-2)},&m' \ \
{\rm is\ \ odd}\end{matrix}\right.
\end{eqnarray*}
which implies that if $\bar{G}(0)=\bar{G}(1)=0$, then
$$\bar{G}(m')=0, \ \ m'\in Z$$

In this way, we may let
$$\bar{G}(0)=\sum_{n=-\infty}^{\infty}(-16\pi^2n^2k\omega-96u_0\pi^2k^2n^2+256\pi^4n^4k^4+c)e^{2\pi
in^2\tau}=0,\eqno(2.7)$$
$$\bar{G}(1)=\sum_{n=-\infty}^{\infty}(-4\pi^2(2n-1)^2k\omega-24u_0\pi^2(2n-1)^2k^2
+16\pi^4(2n-1)^4k^4+c)e^{\pi i(2n^2-2n+1)\tau}=0,\eqno(2.8)$$

Denote
$$\delta_1(n)=e^{2\pi in^2\tau},\ \ \delta_2(n)=e^{\pi i(2n^2-2n+1)\tau},$$
$$a_{11}=-\sum_{n=-\infty}^{\infty}16\pi^2n^2k\delta_1(n),\ \
a_{12}=\sum_{n=-\infty}^{\infty}\delta_1(n),$$
 $$b_1=\sum_{n=-\infty}^{\infty}(256\pi^4n^4k^4-96u_0\pi^2k^2n^2)\delta_1(n),\ \
a_{21}=-\sum_{n=-\infty}^{\infty}4\pi^2(2n-1)^2k\delta_2(n),$$
$$a_{22}=\sum_{n=-\infty}^{\infty}\delta_2(n),\ \
b_2=\sum_{n=-\infty}^{\infty}[16\pi^4(2n-1)^4k^4-24u_0\pi^2(2n-1)^2k^2]\delta_2(n),$$
then (2.7) and (2.8) can be written as
$$a_{11}\omega+a_{12}c+b_1=0,$$
$$a_{21}\omega+a_{22}c+b_2=0.$$
Solving this system, we get
$$\omega=\frac{b_2a_{12}-b_1a_{22}}{a_{11}a_{22}-a_{12}a_{21}},\ \ \ \
c=\frac{b_1a_{21}-b_2a_{11}}{a_{11}a_{22}-a_{12}a_{21}}.\eqno(2.9)$$

Finally we get periodic wave solution
$$u=u_0+2(\ln f)_{xx},\eqno(2.10)$$
where $f$ and  $\omega$ are  given by (2.6) and  (2.9)
respectively. From fig.1-fig.3, we see that the parameter $\tau$
does affect the period and shape of wave, but parameter $k$ has
effect  on the period and shape of wave.

\begin{figure}[h]
$ \ \ \ \ $  {\footnotesize $(a) \ \ \ \ \ \ \ \ \ \ \ \ \ \ \ \ \
\ \ \ \ \ \ \ \ \ \ \ \ \ \ \ \ \ \ \ \ \ \ \ \  $}\
 \ \  \ \ \ \ \ \ \
 \ \ {\footnotesize $(b)$}$ \ \ \ \ \ \ \ \ \ \ \ \  \ \ \ \ \ \ \ \ \ \ \ \ \ \ \ \ \ \ \ \ \ \ \ \ \ \ \
 \ \ \ \ $
 \\[6pt]
\begin{minipage}[u]{0.48\linewidth}
\centering \epsfig{figure=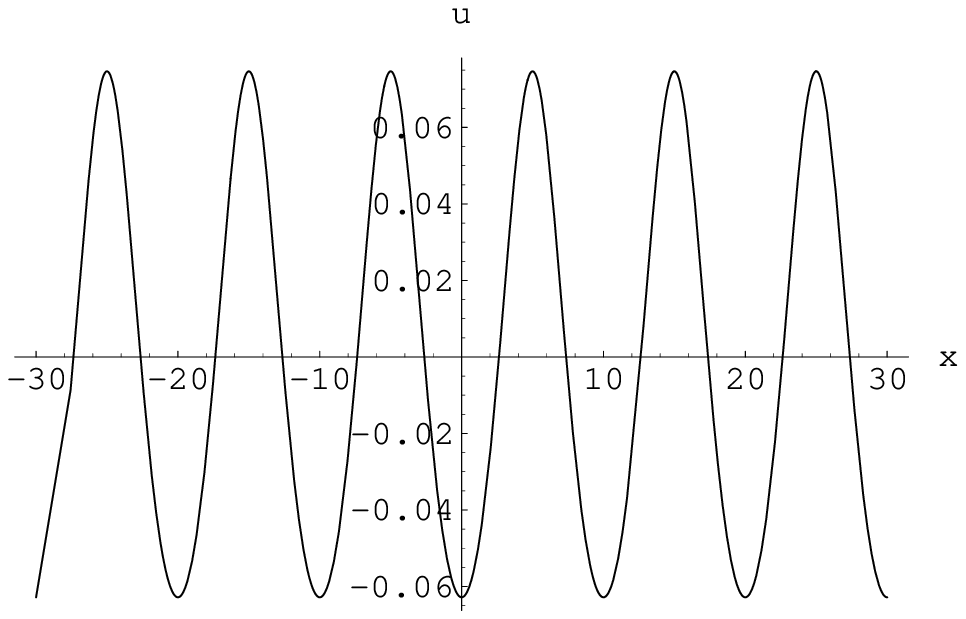,width=2.8 in }
\end{minipage}
\begin{minipage}[v]{0.48\linewidth}
\centering \epsfig{file=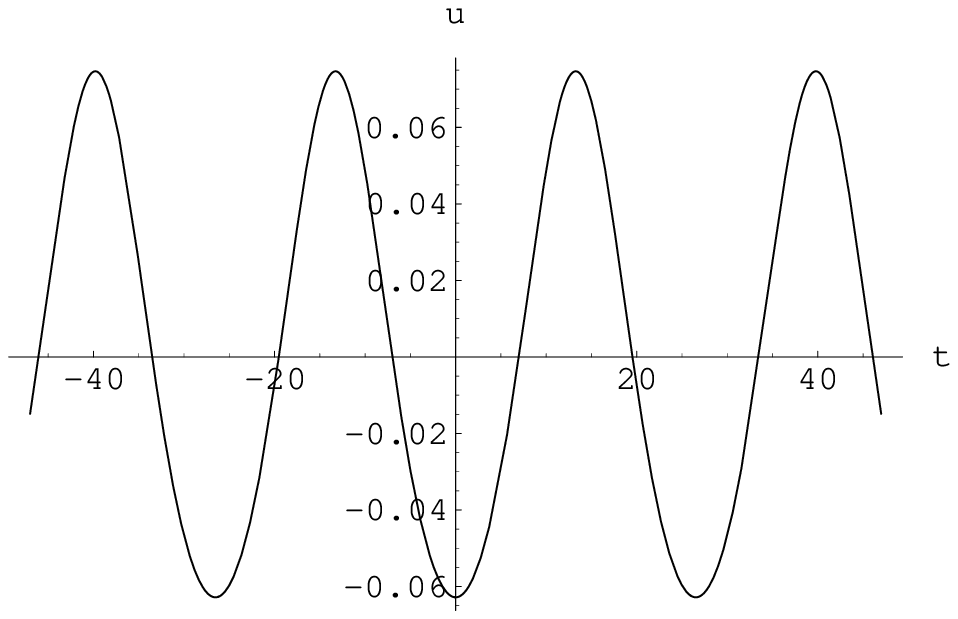,width=2.8 in}
\end{minipage}
\begin{center}
\vspace{6mm}
 {\footnotesize Figure 1. One-periodic wave
for KdV equation and the effect of parameters on wave shape:\\ (a)
along x-axis,  (b) along t-axis, where $k=0.1, \tau=i$.}
\end{center}
\end{figure}

\begin{figure}[h]
$ \ \ \ \ $  {\footnotesize $(a) \ \ \ \ \ \ \ \ \ \ \ \ \ \ \ \ \
\ \ \ \ \ \ \ \ \ \ \ \ \ \ \ \ \ \ \ \ \ \ \ \  $}\
 \ \  \ \ \ \ \ \ \
 \ \ {\footnotesize $(b)$}$ \ \ \ \ \ \ \ \ \ \ \ \  \ \ \ \ \ \ \ \ \ \ \ \ \ \ \ \ \ \ \ \ \ \ \ \ \ \ \
 \ \ \ \ $
 \\[6pt]
\begin{minipage}[u]{0.48\linewidth}
\centering \epsfig{figure=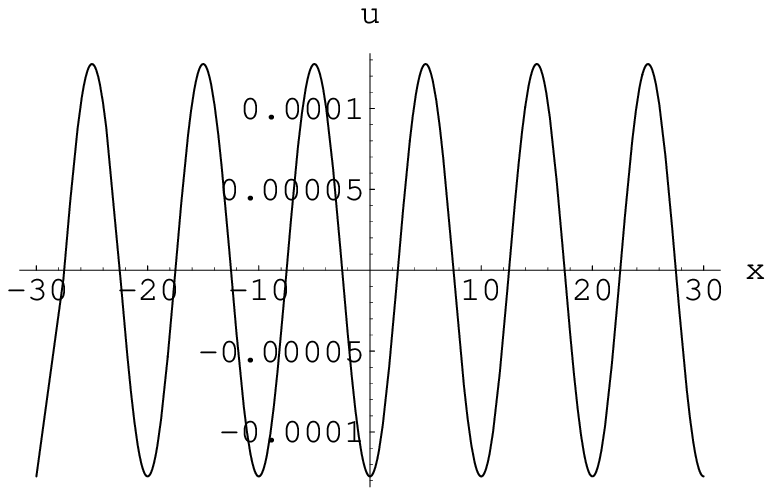,width=2.8 in }
\end{minipage}
\begin{minipage}[v]{0.48\linewidth}
\centering \epsfig{file=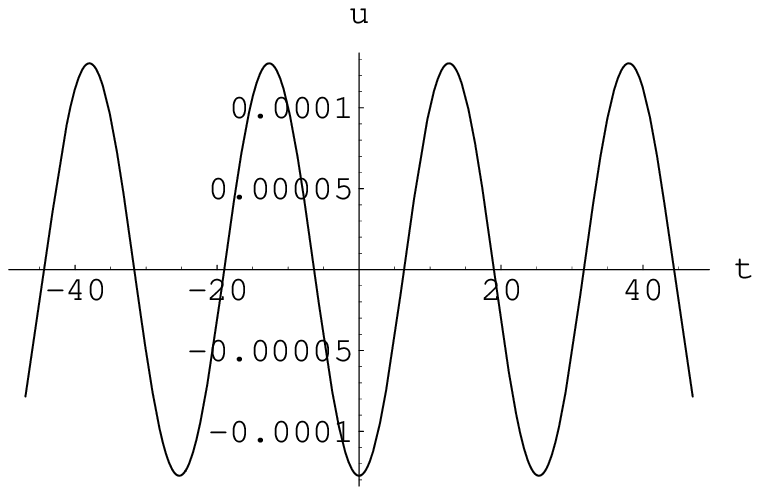,width=2.8 in}
\end{minipage}
\begin{center}
\vspace{6mm}
 {\footnotesize Figure 2. One-periodic wave
for KdV equation and the effect of parameters on wave shape:\\ (a)
along x-axis,  (b) along t-axis, where $k=0.1, \tau=3i$.}
\end{center}
\end{figure}

\begin{figure}[h]
$ \ \ \ \ $  {\footnotesize $(a) \ \ \ \ \ \ \ \ \ \ \ \ \ \ \ \ \
\ \ \ \ \ \ \ \ \ \ \ \ \ \ \ \ \ \ \ \ \ \ \ \  $}\
 \ \  \ \ \ \ \ \ \
 \ \ {\footnotesize $(b)$}$ \ \ \ \ \ \ \ \ \ \ \ \  \ \ \ \ \ \ \ \ \ \ \ \ \ \ \ \ \ \ \ \ \ \ \ \ \ \ \
 \ \ \ \ $
 \\[6pt]
\begin{minipage}[u]{0.48\linewidth}
\centering \epsfig{figure=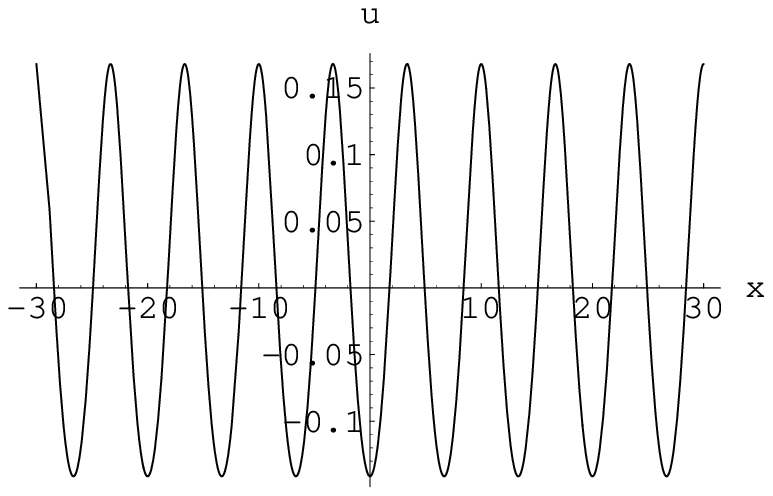,width=2.8 in }
\end{minipage}
\begin{minipage}[v]{0.48\linewidth}
\centering \epsfig{file=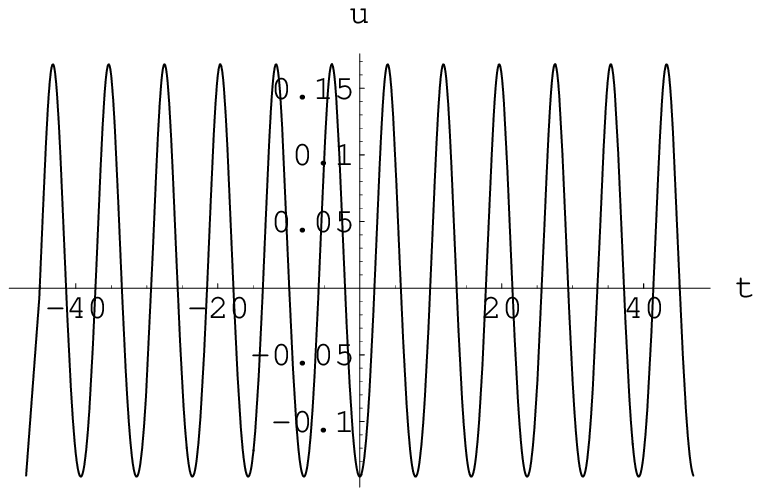,width=2.8 in}
\end{minipage}
\begin{center}
\vspace{6mm}
 {\footnotesize Figure 3. One-periodic wave
for KdV equation and the effect of parameters on wave shape:\\ (a)
along x-axis,  (b) along t-axis, where $k=0.15, \tau=i$.}
\end{center}
\end{figure}

 The well-known soliton solution of KdV equation can
be obtained as limit of the periodic solution (2.10). For this
purpose, we write $f$ as
$$ f=1+\alpha(e^{2\pi i\xi}+e^{-2\pi i\xi})+\alpha^4(e^{4\pi i\xi}+e^{-4\pi i\xi})
+\cdots,$$ where $\alpha=e^{i\pi \tau}$.

 Setting $u_0=0,\ \ \xi=\xi'-\tau/2,\ \ k'=2\pi i k,\ \ \omega'=2\pi i
 \omega$,
 we get
\begin{eqnarray*}
&&f=1+e^{k'x+\omega' t}+\alpha^2e^{-2\pi
 i\xi'}+\alpha^2e^{4\pi i\xi'}+\alpha^6e^{-4\pi i\xi'}+\cdots\\
&&\longrightarrow 1+e^{k'x+\omega' t},\ \ {\rm as}\ \
\alpha\longrightarrow 0
 \end{eqnarray*}

So the periodic solution (2.10) convergence to well-known soliton
solution
$$u=2(\ln f)_{xx}, \ \ f=1+e^{k'x+\omega' t},$$
We only need to prove that
$$\omega'\longrightarrow -k'^3,\ \  {\rm as }\ \ \alpha\longrightarrow
0.\eqno(2.11)$$

 In fact, it is easy to see that
\begin{eqnarray*}
&&a_{11}=-32k\pi^2(\alpha^2+4\alpha^8+\cdots),\\
&&a_{12}=1+2\alpha^2+\cdots,\\
&&a_{22}=2\alpha+\alpha^5+\cdots,\\
&&a_{21}=-8k\pi^2(\alpha+9\alpha^5+\cdots),\\
&&b_1=256k^4\pi^4(\alpha^2+16\alpha^8+\cdots),\\
&&b_2=16k^4\pi^4(2\alpha+81\alpha^5),
 \end{eqnarray*}
 which lead to
 $$b_2a_{12}-b_1a_{22}=32k^4\pi^4\alpha+o(\alpha),\ \ \
 a_{11}a_{22}-a_{12}a_{21}=8k\pi^2\alpha+o(\alpha)$$
 Therefore we have
  $$\omega\longrightarrow 4k^3\pi^2,\  \ \ {\rm as }\ \ \alpha\longrightarrow
0,$$ which implies (2.11).

Now we consider two-periodic wave solution of KdV equation
($N=2$). Substituting (2.5) into (2.3), we have
\begin{eqnarray*}
&&Gf\cdot f=\sum_{m,n\in Z^2} G(D_x,D_t)e^{2\pi i<\xi,n>+\pi
i<\tau n,n>}\cdot e^{2\pi i<\xi,m>+\pi
i<\tau m,m>}\\
&&=\sum_{m,n\in Z^2}G(2\pi
i<n-m,k>,2\pi i<n-m,\omega>)e^{2\pi i<\xi,n+m>+\pi i (<\tau m,m>+<\tau n, n>)}\\
&&\stackrel{n+m=m'}{=}\sum_{m'\in
Z^2}\sum_{n_1,n_2=-\infty}^{\infty}G(2\pi i<2n-m',k>,2\pi
i<2n-m',\omega>)\\
&&\ \ \ \ \ \ \ \ \ \ \ \ \ \times \exp(\pi i(<\tau (n-m'),n-m'>+<\tau n, n>))
\exp(2\pi i<\xi,m'>)\\
&&\equiv\sum_{m'\in Z^2}\bar{G}(m_1', m_2') e^{2\pi i <\xi,m'>}=0.
\end{eqnarray*}

It is easy to calculate that
\begin{eqnarray*}
&&\bar{G}(m_1', m_2')\\
&&=\sum_{n_1,n_2=-\infty}^{\infty}G(2\pi i<2n-m',k>,2\pi
i<2n-m',\omega>)e^{\pi i(<\tau (n-m'),n-m'>+<\tau n, n>)}\\
&&\stackrel{n_j=n_j'+\delta_{jl},
l=1,2}{=}\sum_{n_1,n_2=-\infty}^{\infty}G\left(2\pi
i\sum_{j=1}^{2}(2n_j'-(m_j'-2\delta_{jl}))k_j,2\pi
i\sum_{j=1}^{2}(2n_j'-(m_j'-2\delta_{jl}))\omega_j\right)\\
&&\times \exp\left\{\pi
i\sum_{j,k=1}^{2}[(n_j'+\delta_{jl})\tau_{jk}(n_k'+\delta_{kl})+((m_j'-2\delta_{jl}-n_j')+\delta_{jl})
\tau_{jk}(m_k'-2\delta_{kl}-n_k')+\delta_{kl})]\right\}, \\
&&=\left\{\begin{matrix}\bar{G}(m_1'-2,m_2')e^{2\pi
i(m_1'-1)\tau_{11}+2\pi im_2'\tau_{12}},\ \ \ l=1\cr\cr
\bar{G}(m_1',m_2'-2) e^{2\pi i(m_2'-1)\tau_{22}+2\pi
im_1'\tau_{12}},\ \ \ l=2
\end{matrix}\right.
\end{eqnarray*}
which implies that if
$$\bar{G}(0,0)=\bar{G}(0,1)=\bar{G}(1,0)=\bar{G}(1,1)=0,$$
 then $\bar{G}(m_1',m_2')=0$ and (2.2) is an exact solution
of KdV equation.

Denote
 \begin{eqnarray*}
&&a_{j1}=-\sum_{n_1,n_2=-\infty}^{\infty}
4\pi^2<2n-m^j,k>(2n_1-m_1^j)\delta_j(n),\\
&&a_{j2}=-\sum_{n_1,n_2=-\infty}^{\infty}4\pi^2
<2n-m^j,k>(2n_2-m_2^j)\delta_j(n)\\
&&a_{j3}=-\sum_{n_1,n_2=-\infty}^{\infty}24\pi^2
<2n-m^j,k>^2\delta_j(n),\\
&&a_{j4}=\sum_{n_1,n_2=-\infty}^{\infty}\delta_j(n)\\
&& b_j=-\sum_{n_1,n_2=-\infty}^{\infty}16\pi^4
<2n-m^j,k>^4\delta_j(n),\\
&&\delta_j(n)=e^{\pi i<\tau(n-m^j),n-m^j>+\pi i<\tau n,n>},\\
&&j=1, 2, 3, 4\ \ m^1=(0,0),\ \ m^2=(1,0),\ \ m^3=(0,1),\ \ m^4=(1,1)\\
&&A=(a_{kj})_{4\times 4},\ \ b=(b_1, b_2, b_3, b_4)^T,
\end{eqnarray*}
then we have $$A\left(\begin{matrix}\omega_1\cr\omega_2\cr u_0\cr
c\end{matrix}\right)=b,$$
 from which we obtain
$$\omega_1=\frac{\Delta_1}{\Delta},\ \ \omega_2=\frac{\Delta_2}{\Delta},\ \
 u_0=\frac{\Delta_3}{\Delta},\eqno(2.12)$$
where $\Delta=|A|$ and $\Delta_1, \Delta_2, \Delta_3$ are produced
from $\Delta$ by replacing its $1th, 2th, 3th$ column with $b$
respectively. Finally we get two-periodic wave solution (See
fig.2)
$$u=u_0+2(\ln f)_{xx},\eqno(2.13)$$
where $f$ and $\omega_1, \omega_2$  are given by (2.5) and (2.12)
respectively. The properties of the solution are shown in
fig.4-fig.7. Fig.4 and fig.5 shown effect of parameters $k_1, k_2$
on period and shape of wave.  Fig.7 gives effect of parameter
$\tau_{11}, \tau_{12}, \tau_{22}$ on period and shape of wave. The
fig.7 and fig.8 show that two-periodic wave can degenerate to
one-periodic wave when $k_1$ is sufficient small.

\begin{figure}[h]
$ \ \ \ \ $  {\footnotesize $(a) \ \ \ \ \ \ \ \ \ \ \ \ \ \ \ \ \
\ \ \ \ \ \ \ \ \ \ \ \ \ \ \ \ \ \ \ \ \ \ \ \  $}\
 \ \  \ \ \ \ \ \ \
 \ \ {\footnotesize $(b)$}$ \ \ \ \ \ \ \ \ \ \ \ \  \ \ \ \ \ \ \ \ \ \ \ \ \ \ \ \ \ \ \ \ \ \ \ \ \ \ \
 \ \ \ \ $
 \\[6pt]
\begin{minipage}[u]{0.48\linewidth}
\centering \epsfig{figure=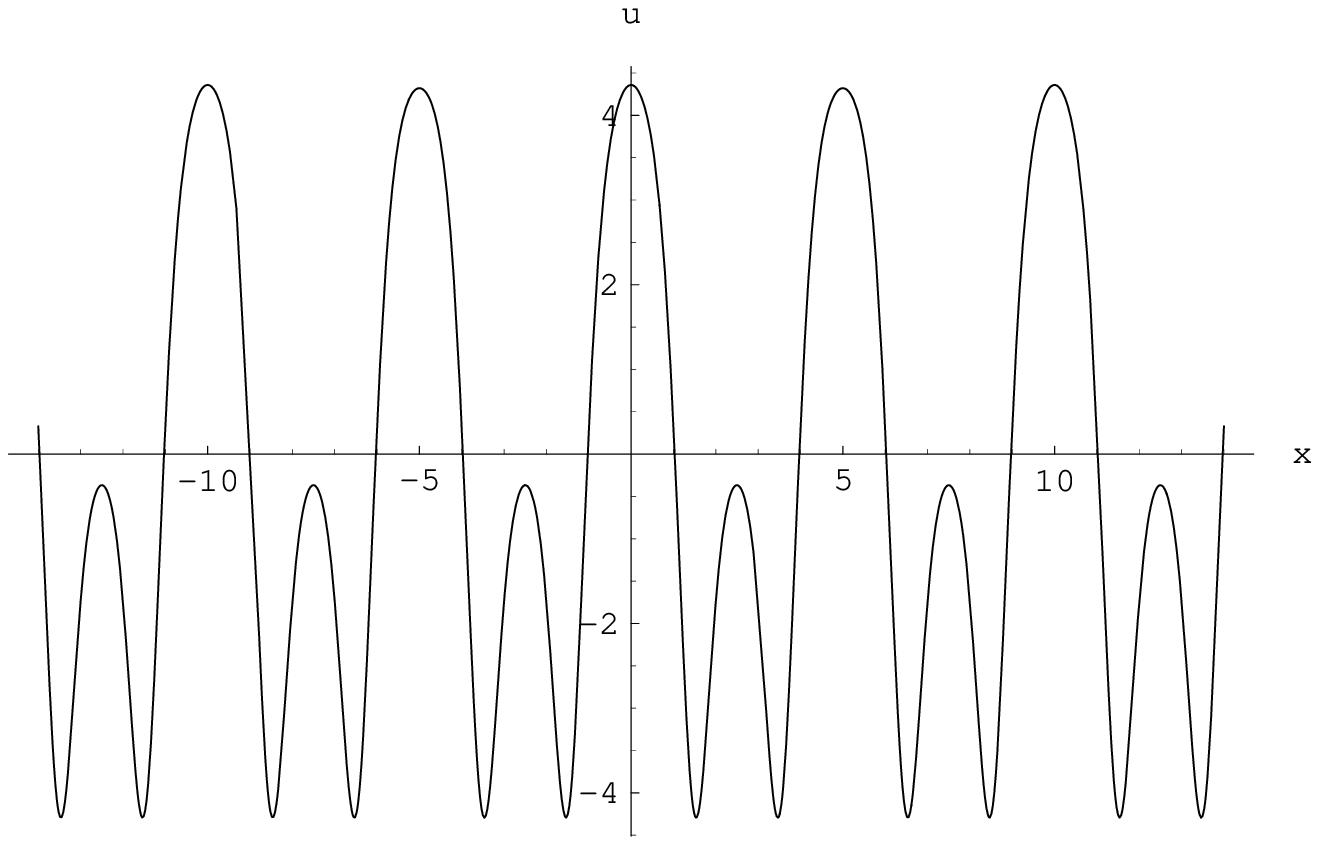,width=2.8 in }
\end{minipage}
\begin{minipage}[v]{0.48\linewidth}
\centering \epsfig{file=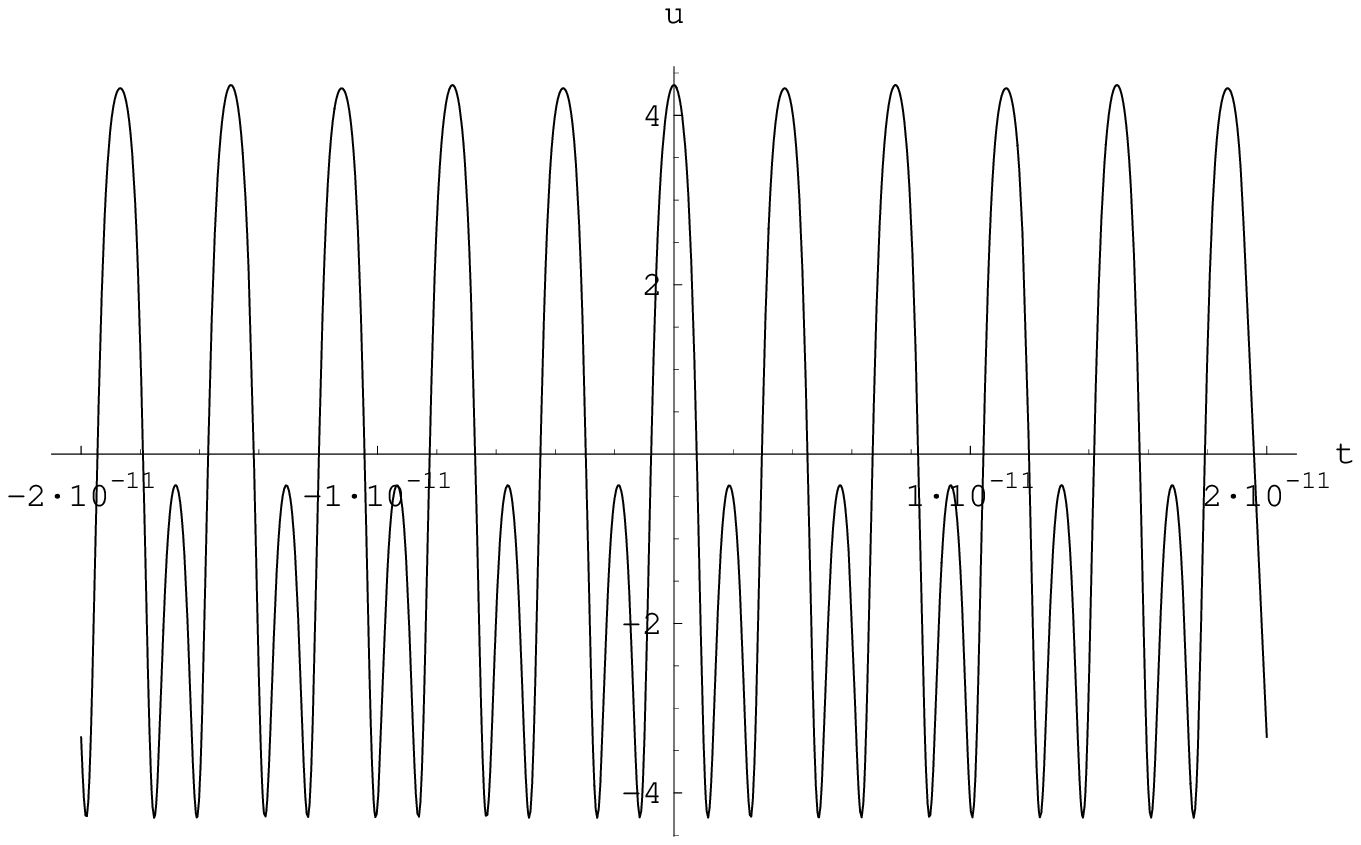,width=2.8 in}
\end{minipage}
\begin{center}
\vspace{6mm}
 {\footnotesize Figure 4. Two-periodic wave
for KdV equation and the effect of parameters on wave shape:\\
(a) along x-axis, (b) along t-axis, where $\tau_{11}=0.1i,
\tau_{12}=0.2i, \tau_{22}=3i, k_1=0.2, k_2=-0.3$.}
\end{center}
\end{figure}

\begin{figure}[h]
$ \ \ \ \ $  {\footnotesize $(a) \ \ \ \ \ \ \ \ \ \ \ \ \ \ \ \ \
\ \ \ \ \ \ \ \ \ \ \ \ \ \ \ \ \ \ \ \ \ \ \ \  $}\
 \ \  \ \ \ \ \ \ \
 \ \ {\footnotesize $(b)$}$ \ \ \ \ \ \ \ \ \ \ \ \  \ \ \ \ \ \ \ \ \ \ \ \ \ \ \ \ \ \ \ \ \ \ \ \ \ \ \
 \ \ \ \ $
 \\[6pt]
\begin{minipage}[u]{0.48\linewidth}
\centering \epsfig{figure=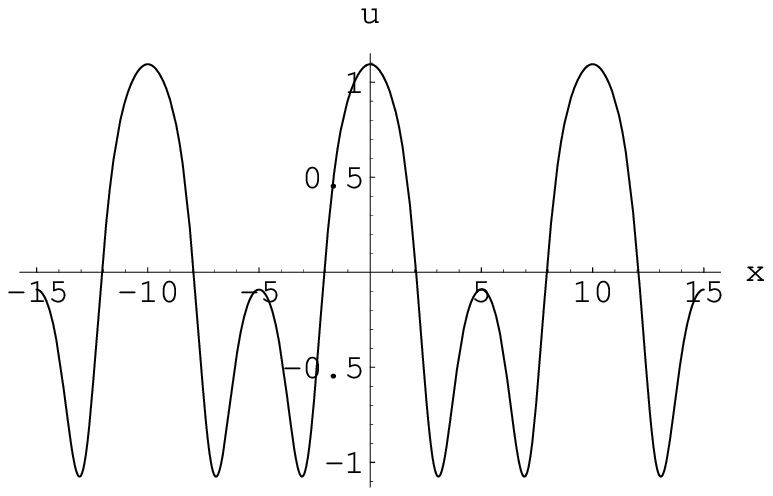,width=2.8 in }
\end{minipage}
\begin{minipage}[v]{0.48\linewidth}
\centering \epsfig{file=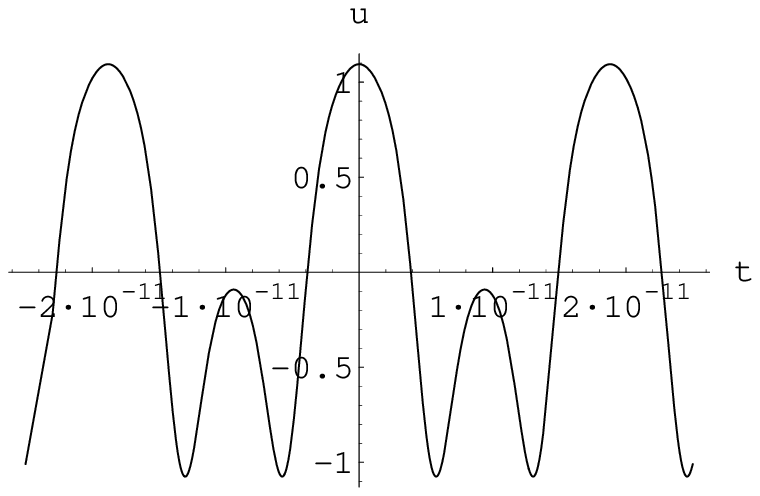,width=2.8 in}
\end{minipage}
\begin{center}
\vspace{6mm}
 {\footnotesize Figure 5. Two-periodic wave
for KdV equation and the effect of parameters on wave shape:\\ (a)
along x-axis,  (b) along t-axis, where $\tau_{11}=0.1i,
\tau_{12}=0.2i, \tau_{22}=3i, k_1=0.1, k_2=-0.3$.}
\end{center}
\end{figure}

\begin{figure}[h]
$ \ \ \ \ $  {\footnotesize $(a) \ \ \ \ \ \ \ \ \ \ \ \ \ \ \ \ \
\ \ \ \ \ \ \ \ \ \ \ \ \ \ \ \ \ \ \ \ \ \ \ \  $}\
 \ \  \ \ \ \ \ \ \
 \ \ {\footnotesize $(b)$}$ \ \ \ \ \ \ \ \ \ \ \ \  \ \ \ \ \ \ \ \ \ \ \ \ \ \ \ \ \ \ \ \ \ \ \ \ \ \ \
 \ \ \ \ $
 \\[6pt]
\begin{minipage}[u]{0.48\linewidth}
\centering \epsfig{figure=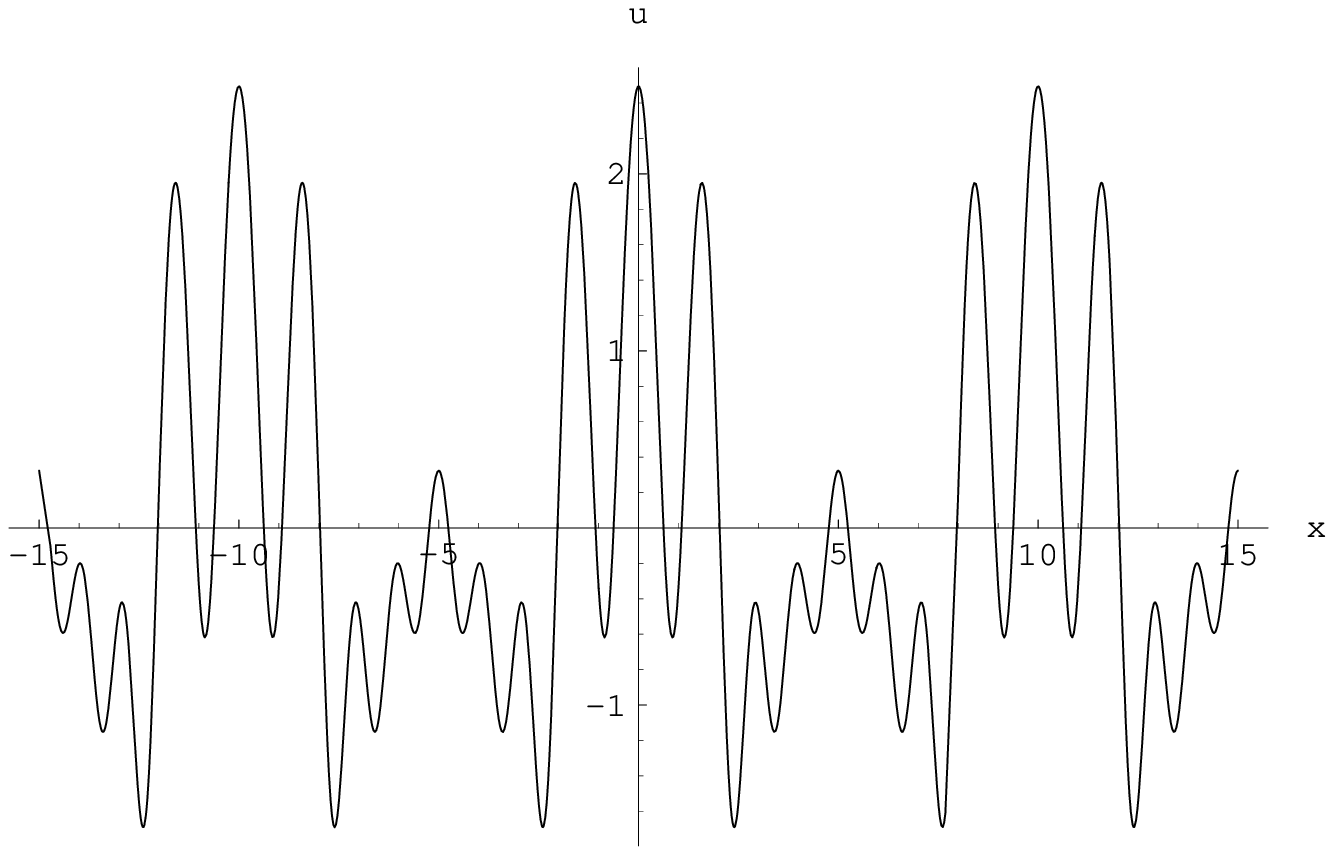,width=2.8 in }
\end{minipage}
\begin{minipage}[v]{0.48\linewidth}
\centering \epsfig{file=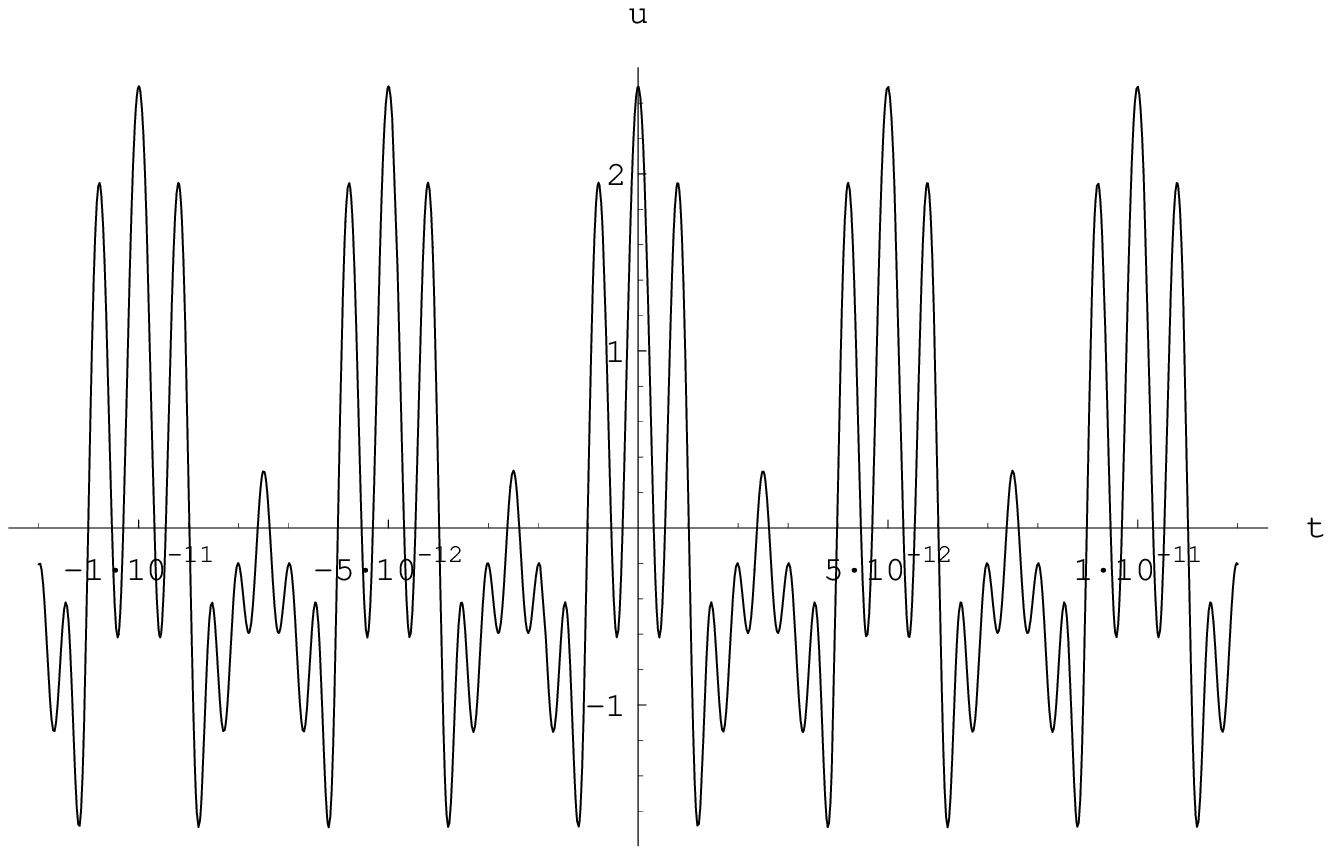,width=2.8 in}
\end{minipage}
\begin{center}
\vspace{6mm}
 {\footnotesize Figure 6. Two-periodic wave
for KdV equation and the effect of parameters on wave shape:\\ (a)
along x-axis,  (b) along t-axis, where $\tau_{11}=0.1,
\tau_{12}=0.3i, \tau_{22}=2i, k_1=0.1, k_2=-0.3$.}
\end{center}
\end{figure}

\begin{figure}[h]
$ \ \ \ \ $  {\footnotesize $(a) \ \ \ \ \ \ \ \ \ \ \ \ \ \ \ \ \
\ \ \ \ \ \ \ \ \ \ \ \ \ \ \ \ \ \ \ \ \ \ \ \  $}\
 \ \  \ \ \ \ \ \ \
 \ \ {\footnotesize $(b)$}$ \ \ \ \ \ \ \ \ \ \ \ \  \ \ \ \ \ \ \ \ \ \ \ \ \ \ \ \ \ \ \ \ \ \ \ \ \ \ \
 \ \ \ \ $
 \\[6pt]
\begin{minipage}[u]{0.48\linewidth}
\centering \epsfig{figure=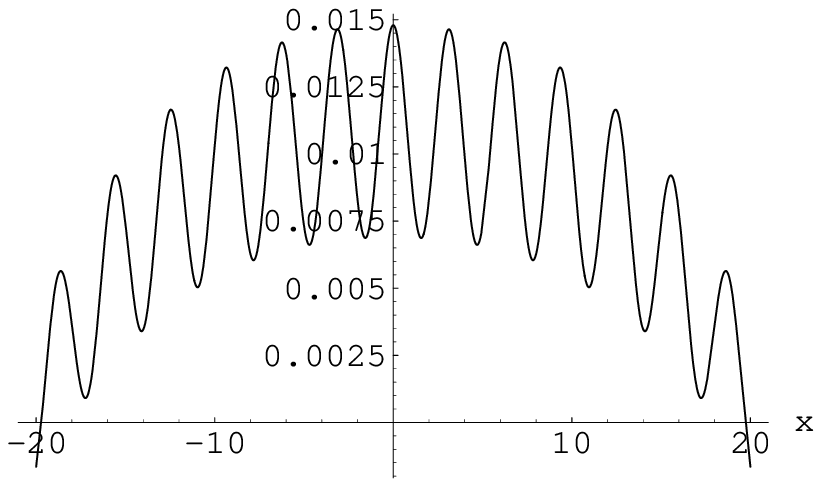,width=2.8 in }
\end{minipage}
\begin{minipage}[v]{0.48\linewidth}
\centering \epsfig{file=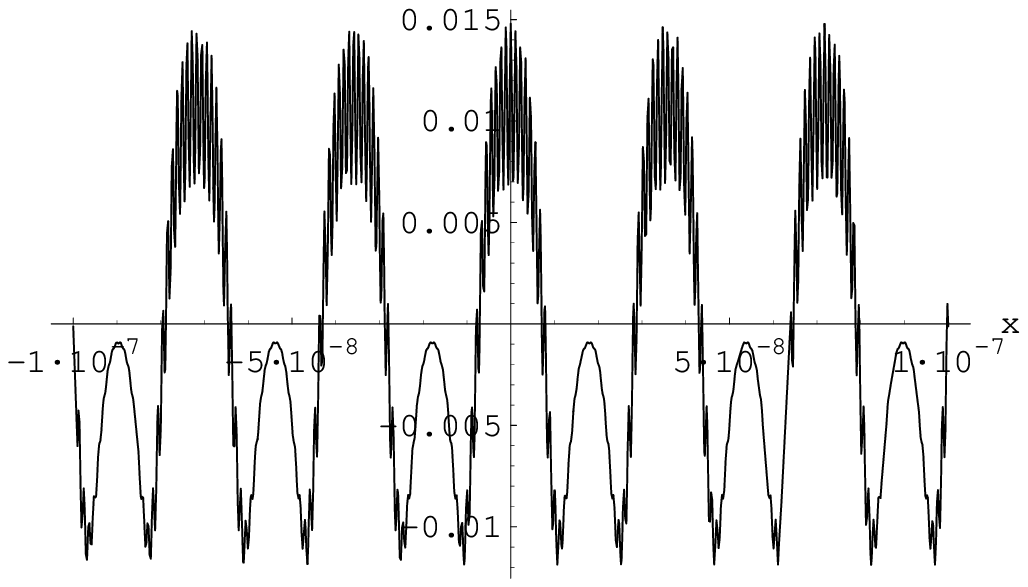,width=2.8 in}
\end{minipage}
\begin{center}
\vspace{6mm}
 {\footnotesize Figure 7. Two-periodic wave
for KdV equation and the effect of parameters on wave shape:\\ (a)
along x-axis,  (b) along t-axis, where $\tau_{11}=0.01,
\tau_{12}=0.2i, \tau_{22}=3i, k_1=0.1, k_2=-0.2$.}
\end{center}
\end{figure}

\begin{figure}[h]
$ \ \ \ \ $  {\footnotesize $(a) \ \ \ \ \ \ \ \ \ \ \ \ \ \ \ \ \
\ \ \ \ \ \ \ \ \ \ \ \ \ \ \ \ \ \ \ \ \ \ \ \  $}\
 \ \  \ \ \ \ \ \ \
 \ \ {\footnotesize $(b)$}$ \ \ \ \ \ \ \ \ \ \ \ \  \ \ \ \ \ \ \ \ \ \ \ \ \ \ \ \ \ \ \ \ \ \ \ \ \ \ \
 \ \ \ \ $
 \\[6pt]
\begin{minipage}[u]{0.48\linewidth}
\centering \epsfig{figure=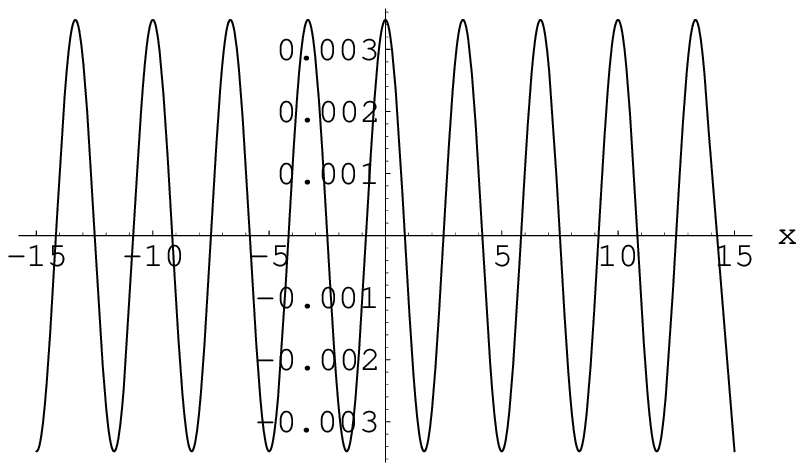,width=2.8 in }
\end{minipage}
\begin{minipage}[v]{0.48\linewidth}
\centering \epsfig{file=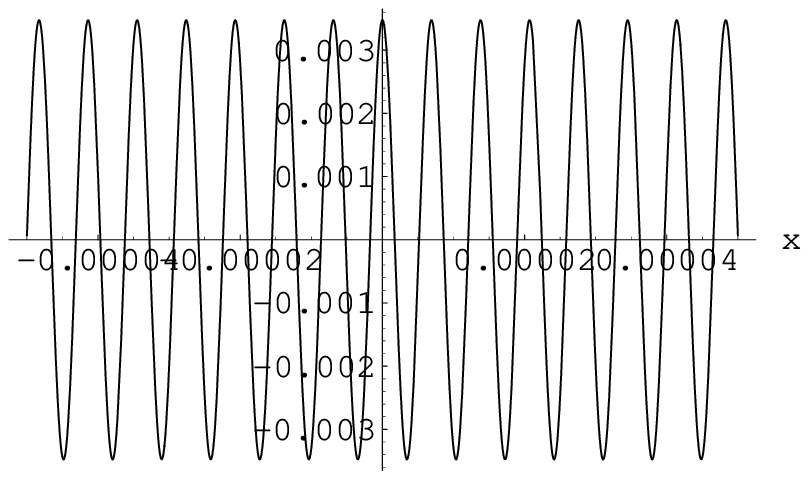,width=2.8 in}
\end{minipage}
\begin{center}
\vspace{6mm}
 {\footnotesize Figure 8. Two-periodic wave
for KdV equation and the effect of parameters on wave shape:\\ (a)
along x-axis,  (b) along t-axis, where $\tau_{11}=0.001,
\tau_{12}=0.2i, \tau_{22}=3i, k_1=0.1, k_2=-0.2$.}
\end{center}
\end{figure}

The two-soliton solution of KdV equation can be obtained as limit
of the periodic solution (2.10).  We write $f$ as
\begin{eqnarray*}
&& f=1+(e^{2\pi i\xi_1}+e^{-2\pi i\xi_1})e^{\pi i\tau_{11}}
+(e^{2\pi i\xi_2}+e^{-2\pi i\xi_2})e^{\pi
i\tau_{22}}\\
&&+(e^{2\pi i(\xi_1+\xi_2)}+e^{-2\pi i(\xi_1+\xi_2)})e^{\pi
i(\tau_{11}+2\tau_{12}+\tau_{22})}+\cdots
\end{eqnarray*}

 Setting $ \xi_1'=2\pi i\xi_1-\pi i\tau_{11},\ \ \xi_2'=2\pi i\xi_2-\pi i\tau_{22},
 \ \ \tau_{12}=i\widetilde{\tau}$ ($\widetilde{\tau}$ is a real),
 we get
\begin{eqnarray*}
&&f=1+e^{\xi_1'}+e^{\xi_2'}+e^{\xi_1'+\xi_2'+2\pi i\tau_{12}
}+\alpha_1^2e^{-\xi_1'}
 +\alpha_2^2e^{-\xi_2'}+\alpha_1^2\alpha_2^2e^{-\xi_1-\xi_2+2\pi i\tau_{11}}+\cdots\\
&&\longrightarrow 1+e^{\xi_1'}+e^{\xi_2'}+e^{\xi_1'+\xi_2'-2\pi
\widetilde{\tau}},\ \ {\rm as}\ \ \alpha_1, \alpha_2
\longrightarrow 0
 \end{eqnarray*}
where
$$\alpha_1=e^{\pi i\tau_{11}},\ \ \alpha_2=e^{\pi
i\tau_{22}}, \ \ \xi_i'=k_i'x+\omega_i't,\  \ \ e^{-2\pi
\widetilde{\tau}}=\left(\frac{k_1'-k_2'}{\omega_1'-\omega_2'}\right)^2,$$
$$\omega_i'\rightarrow -k_i'^3,\ \ {\rm as}\ \ \alpha_1,
\alpha_2\rightarrow0.$$
\\[24pt]
{\bf 3.  Periodic wave solutions of KP equation and their
reduction}

Consider KP equation
$$u_t=(u_{xx}+3u^2)_x+3\partial^{-1}u_{yy}.\eqno(3.1)$$
Substituting transformation
$$u=u_0+2(\ln f)_{xx}\eqno(3.2)$$
into (3.1) and integrating once again, we then get the following
bilinear form
$$G(D_x,D_y,D_t)f\cdot f=(D_xD_t-6u_0D_x^2-D_x^4-3D_y^2+c_0)f\cdot f=0\eqno(3.3)$$
where $c_0$ is a constant of integration  and $u_0$ is a constant
solution of KP equation. Let us consider Riemann theta  function
solution of KP equation
$$f=\sum_{n\in Z^N}e^{\pi i<\tau n,n>+2\pi i<\xi,n>}\eqno(3.4)$$
where $n=(n_1,\cdots,n_N), \xi =(\xi_1, \cdots, \xi_N)$, $\tau$ is
a symmetric matrix and ${\rm Im}\tau>0$,
$\xi_j=k_j(x+\rho_jy+\omega_j t)+\gamma_j, j=1,\cdots, N.$

In the following, we consider some special cases. When $N=1$,
(3.4) is reduced to
$$f=\sum_{n=-\infty}^{\infty}e^{2\pi in\xi+\pi in^2\tau}\eqno(3.5)$$
Substituting (3.5) into (3.3) gives
{\footnotesize\begin{eqnarray*} &&Gf\cdot
f=G(D_x,D_y,D_t)\sum_{n=-\infty}^{\infty}e^{2\pi in\xi+\pi
in^2\tau}\cdot\sum_{m=-\infty}^{\infty}
e^{2\pi i m\xi+\pi i m^2\tau}\\
&&=\sum_{n=-\infty}^{\infty}\sum_{m=-\infty}^{\infty}G(D_x,D_y,D_t)e^{2\pi
in\xi+\pi in^2\tau}\cdot e^{2\pi i m\xi+\pi i
m^2\tau}\\
&&=\sum_{n=-\infty}^{\infty}\sum_{m=-\infty}^{\infty}G(2\pi
i(n-m)k,2\pi i(n-m)k\rho,2\pi i(n-m)k\omega)e^{2\pi i(n+m)\xi+\pi i ( n^2+m^2)\tau}\\
&&\stackrel{n+m=m'}{=}\sum_{m'=-\infty}^{\infty}\left\{
\sum_{n=-\infty}^{\infty}G(2\pi i(2n-m')k,2\pi i(2n-m')k\rho,2\pi
i(2n-m')k\omega)e^{\pi
i[(n^2+(n-m')^2]\tau}\right\}e^{2\pi i m'\xi}\\
&&\equiv\sum_{m'=-\infty}^{\infty}\bar{G}(m')e^{2\pi i m'\xi}=0.
\end{eqnarray*}}
Noticing that
\begin{eqnarray*}
&&\bar{G}(m')=\sum_{n=-\infty}^{\infty}G(2\pi i(2n-m')k,2\pi
i(2n-m')k\rho,2\pi
i(2n-m')k\omega)e^{\pi i[(n^2+(n-m')^2]\tau}\\
&&\stackrel{n=n'+1}{=}\sum_{n'=-\infty}^{\infty}G(2\pi
i[2n'-(m'-2)]k,2\pi i[2n'-(m'-2)]k\rho,2\pi i[2n'-(m'-2)]k\omega)\\
&&\ \ \  \  \ \times \exp(\pi i[(n'^2+(n'-(m'-2))^2]\tau)
\exp(2\pi i(m'-1)\tau)\\
&&=\bar{G}(m'-2)e^{2\pi i (m'-1)\tau}\\
&&=\cdots=\left\{\begin{matrix}\bar{G}(0)e^{\pi im'(m'-1)},&m' \ \
{\rm is\ \ even}\cr\cr \bar{G}(1)e^{\pi i(m'+1)(m'-2)},&m' \ \
{\rm is\ \ odd}\end{matrix}\right.
\end{eqnarray*}
which implies that if $\bar{G}(0)=\bar{G}(1)=0$, then
$$\bar{G}(m')=0, \ \ m'\in Z$$
In this way, we may let
$$\bar{G}(0)=\sum_{n=-\infty}^{\infty}(-16\pi^2n^2k^2\omega+96u_0\pi^2k^2n^2-256\pi^4n^4k^4
+48\pi^2k^2\rho^2n^2+c_0)e^{2\pi in^2\tau}=0,\eqno(3.6)$$
$$\bar{G}(1)=\sum_{n=-\infty}^{\infty}(-4\pi^2(2n-1)^2k^2\omega+24u_0\pi^2(2n-1)^2k^2
-16\pi^4(2n-1)^4k^4$$
$$+12\pi^2(2n-1)^2k^2\rho^2+c_0)e^{\pi
i(2n^2-2n+1)\tau}=0,\eqno(3.7)$$ Denote
$$\alpha=e^{\pi\tau i},\quad a_{11}=-16\pi^2k^2\sum_{n=-\infty}^{\infty}n^2\alpha^{2n^2},\ \
a_{12}=\sum_{n=-\infty}^{\infty}\alpha^{2n^2},$$ {\footnotesize
$$b_1=-\sum_{n=-\infty}^{\infty}(96u_0\pi^2n^2k^2-256\pi^4n^4k^4+48\pi^2k^2\rho^2n^2)\alpha^{2n^2},\
\
a_{21}=-4\pi^2k^2\sum_{n=-\infty}^{\infty}(2n-1)^2\alpha^{2n^2-2n+1},$$
$$a_{22}=\sum_{n=-\infty}^{\infty}\alpha^{2n^2-2n+1},\ \
b_2=\sum_{n=-\infty}^{\infty}[16\pi^4(2n-1)^4k^4-24u_0\pi^2(2n-1)^2k^2-12\pi^2(2n-1)^2k^2\rho^2
]\alpha^{2n^2-2n+1},$$}
then (3.6) and (3.7) can be written as
$$a_{11}\omega+a_{12}c_0=b_1,$$
$$a_{21}\omega+a_{22}c_0=b_2.$$
Solving this system, we get
$$\omega=\frac{b_1a_{22}-b_2a_{12}}{a_{11}a_{22}-a_{12}a_{21}},\ \ \ \
c_0=\frac{b_2a_{11}-b_1a_{21}}{a_{11}a_{22}-a_{12}a_{21}}.\eqno(3.8)$$
Finally we get periodic wave solution of KP equation (3.1)
$$u=u_0+2(\ln f)_{xx},\eqno(3.9)$$
where $f$ and  $\omega$ are  given by (3.4) and  (3.8)
respectively. From fig.9-fig.11, we see that the parameter $\tau$
does affect the period and shape of wave, but parameter $k$ has
effect  on the period and shape of wave.

\begin{figure}[h]
$ \ \ \ \ $  {\footnotesize $ \ \ \ \ \ \ \ \ \ \ \ \ \ \ \ \ \ \
\ $}\
  {\footnotesize $(a)$}$ \ \ \ \ \ \ \ \ \ \ \ \  \ \ \ \ \ \ \ \ \ \ \ \ \
   \ \ \ \ \ \ \ \ \ \ \ \ \ \ \ \ \ \ \ \ \ \ \ \ \ \ \
 \ \ \ \ $ \\[6pt]
\begin{minipage}[u]{0.48\linewidth}
\centering \epsfig{figure=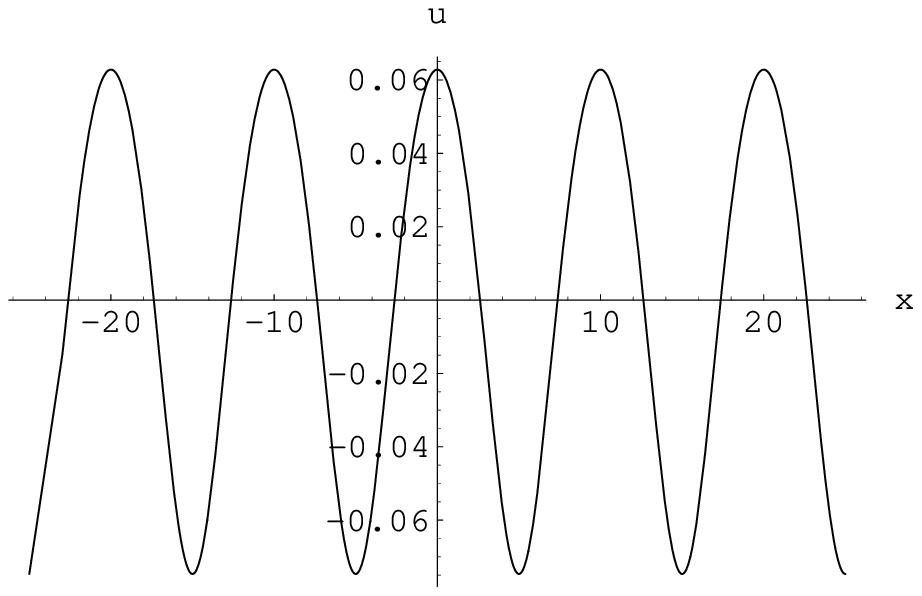,width=2.8 in }
\end{minipage}\\[12pt]
$ \ \ \ \ $  {\footnotesize $(b) \ \ \ \ \ \ \ \ \ \ \ \ \ \ \ \ \
\ \ \ \ \ \ \ \ \ \ \ \ \ \ \ \ \ \ \ \ \ \ \ \  $}\
 \ \  \ \ \ \ \ \ \
 \ \ {\footnotesize $(c)$}$ \ \ \ \ \ \ \ \ \ \ \ \  \ \ \ \ \ \ \ \ \ \ \ \ \ \ \ \ \ \ \ \ \ \ \ \ \ \ \
 \ \ \ \ $ \\[6pt]
\begin{minipage}[u]{0.48\linewidth}
\centering \epsfig{figure=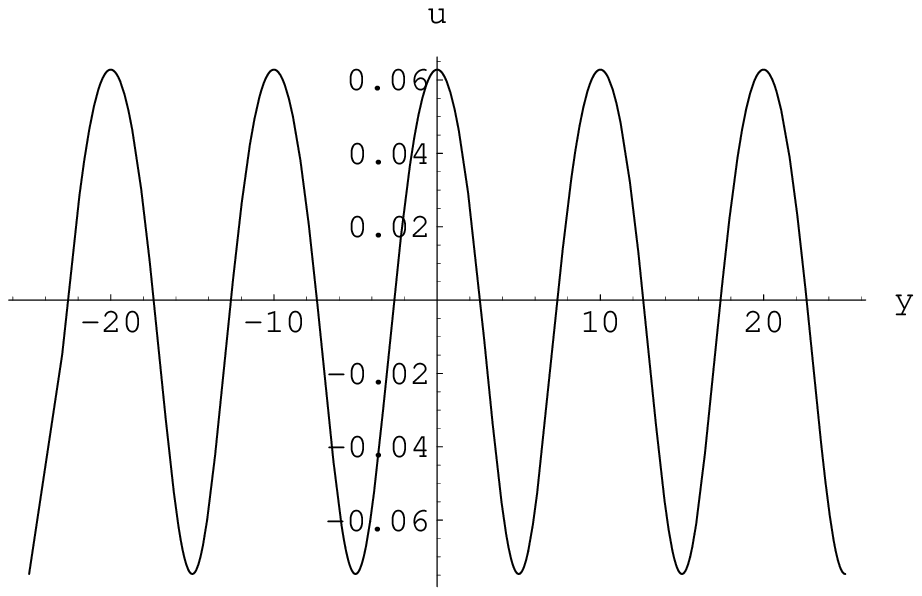,width=2.8 in }
\end{minipage}
\begin{minipage}[v]{0.48\linewidth}
\centering \epsfig{file=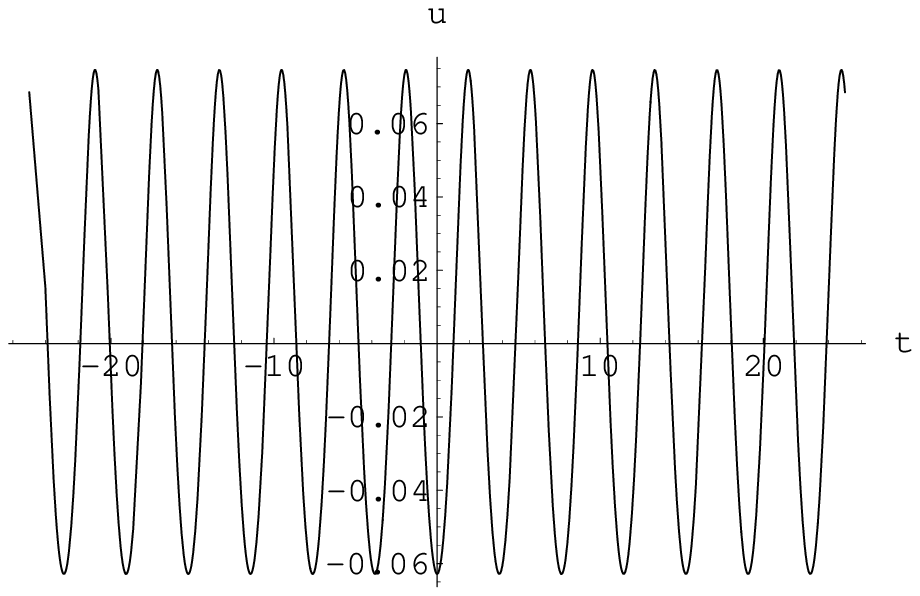,width=2.8 in}
\end{minipage}
\begin{center}
\vspace{6mm}
 {\footnotesize Figure 9. One-periodic wave
for KP equation and the effect of parameters on wave shape:\\
(a) along x-axis,  (b) along y-axis, (b) along t-axis, where
$k=0.1, \tau=i, \rho=2$.}
\end{center}
\end{figure}

\begin{figure}[h]
$ \ \ \ \ $  {\footnotesize $ \ \ \ \ \ \ \ \ \ \ \ \ \ \ \ \ \ \
\ $}\
  {\footnotesize $(a)$}$ \ \ \ \ \ \ \ \ \ \ \ \  \ \ \ \ \ \ \ \ \ \ \ \ \
   \ \ \ \ \ \ \ \ \ \ \ \ \ \ \ \ \ \ \ \ \ \ \ \ \ \ \
 \ \ \ \ $ \\[6pt]
\begin{minipage}[u]{0.48\linewidth}
\centering \epsfig{figure=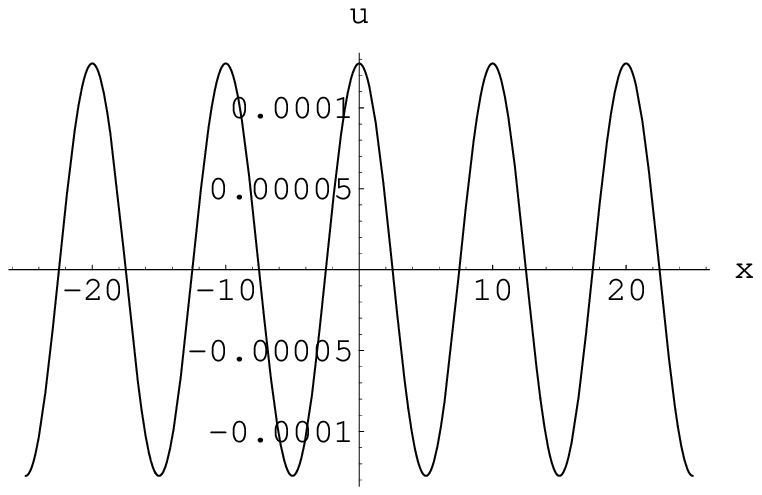,width=2.8 in }
\end{minipage}\\[12pt]
$ \ \ \ \ $  {\footnotesize $(b) \ \ \ \ \ \ \ \ \ \ \ \ \ \ \ \ \
\ \ \ \ \ \ \ \ \ \ \ \ \ \ \ \ \ \ \ \ \ \ \ \  $}\
 \ \  \ \ \ \ \ \ \
 \ \ {\footnotesize $(c)$}$ \ \ \ \ \ \ \ \ \ \ \ \  \ \ \ \ \ \ \ \ \ \ \ \ \ \ \ \ \ \ \ \ \ \ \ \ \ \ \
 \ \ \ \ $ \\[6pt]
\begin{minipage}[u]{0.48\linewidth}
\centering \epsfig{figure=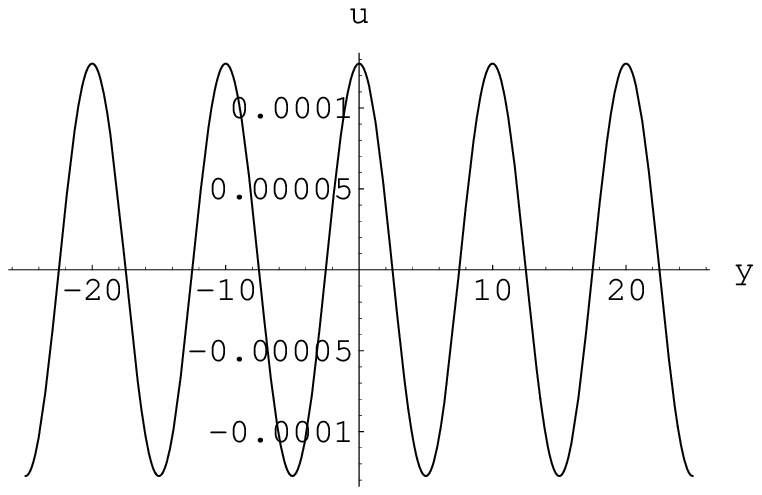,width=2.8 in }
\end{minipage}
\begin{minipage}[v]{0.48\linewidth}
\centering \epsfig{file=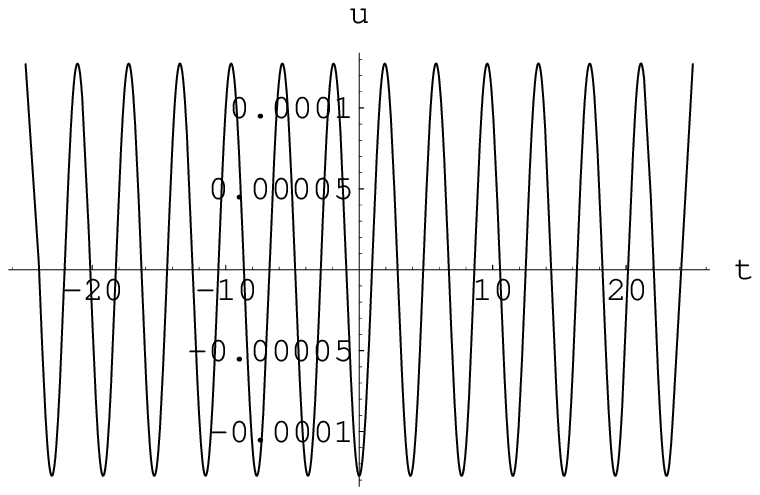,width=2.8 in}
\end{minipage}
\begin{center}
\vspace{6mm}
 {\footnotesize Figure 10. One-periodic wave
for KP equation and the effect of parameters on wave shape:\\
(a) along x-axis,  (b) along y-axis, (b) along t-axis, where
$k=0.1, \tau=3i, \rho=1$.}
\end{center}
\end{figure}

\begin{figure}[h]
$ \ \ \ \ $  {\footnotesize $ \ \ \ \ \ \ \ \ \ \ \ \ \ \ \ \ \ \
\ $}\
  {\footnotesize $(a)$}$ \ \ \ \ \ \ \ \ \ \ \ \  \ \ \ \ \ \ \ \ \ \ \ \ \
   \ \ \ \ \ \ \ \ \ \ \ \ \ \ \ \ \ \ \ \ \ \ \ \ \ \ \
 \ \ \ \ $ \\[6pt]
\begin{minipage}[u]{0.48\linewidth}
\centering \epsfig{figure=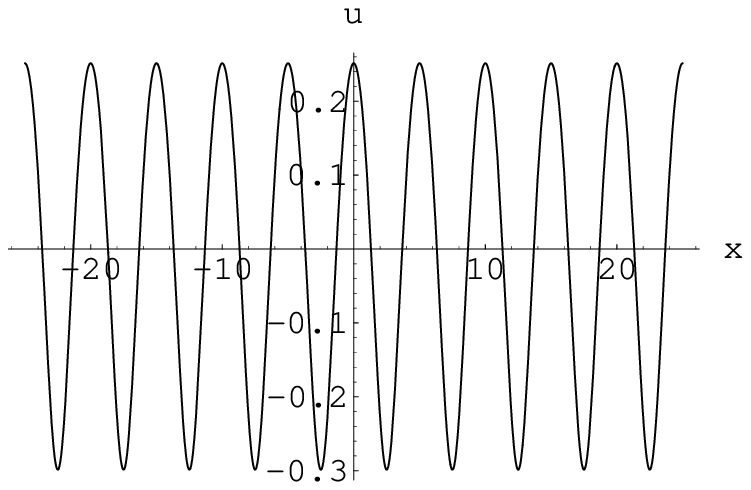,width=2.8 in }
\end{minipage}\\[12pt]
$ \ \ \ \ $  {\footnotesize $(b) \ \ \ \ \ \ \ \ \ \ \ \ \ \ \ \ \
\ \ \ \ \ \ \ \ \ \ \ \ \ \ \ \ \ \ \ \ \ \ \ \  $}\
 \ \  \ \ \ \ \ \ \
 \ \ {\footnotesize $(c)$}$ \ \ \ \ \ \ \ \ \ \ \ \  \ \ \ \ \ \ \ \ \ \ \ \ \ \ \ \ \ \ \ \ \ \ \ \ \ \ \
 \ \ \ \ $ \\[6pt]
\begin{minipage}[u]{0.48\linewidth}
\centering \epsfig{figure=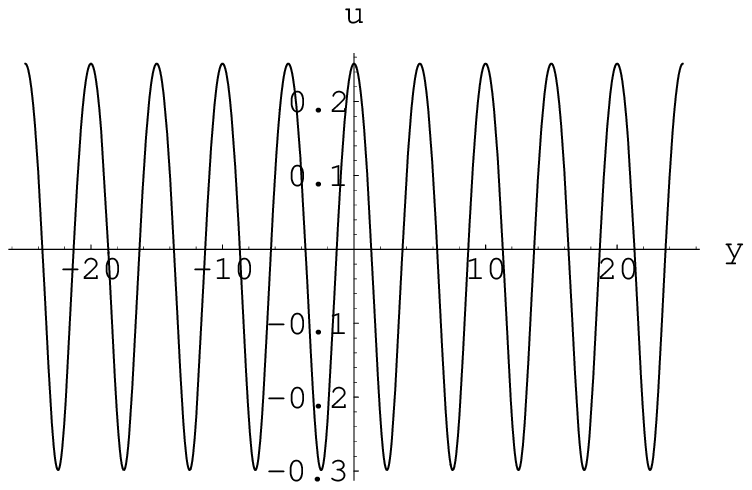,width=2.8 in }
\end{minipage}
\begin{minipage}[v]{0.48\linewidth}
\centering \epsfig{file=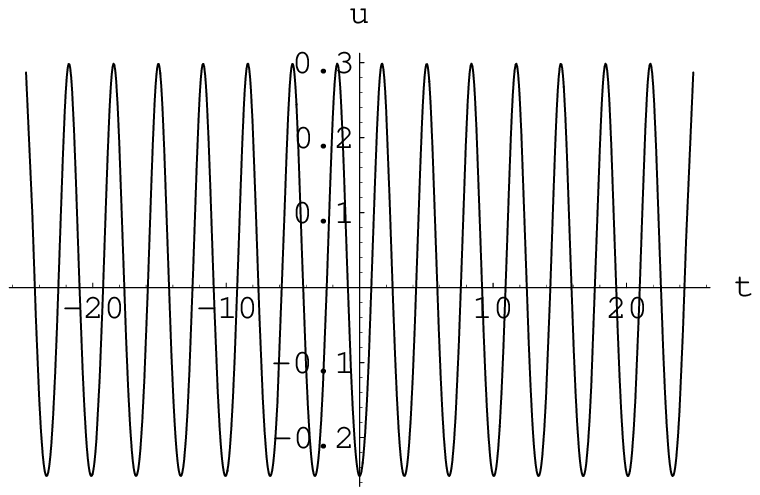,width=2.8 in}
\end{minipage}
\begin{center}
\vspace{6mm}
 {\footnotesize Figure 11. One-periodic wave
for KP equation and the effect of parameters on wave shape:\\
(a) along x-axis,  (b) along y-axis, (c) along t-axis, where
$k=0.2, \tau=i, \rho=1$.}
\end{center}
\end{figure}

The well-known soliton solution of the KP equation can be obtained
as limit of the periodic solution (3.9). To this end, we write $f$
as
$$ f=1+\alpha(e^{2\pi i\xi}+e^{-2\pi i\xi})+\alpha^4(e^{4\pi i\xi}+e^{-4\pi i\xi})
+\cdots .$$ Setting $u_0=0,\ \gamma=\gamma_0-\tau/2,\ k'=2\pi i
k,\ \gamma'_0=2\pi i \gamma_0$,
 we get
\begin{eqnarray*}
&&f=1+e^{k'(x+\rho y+\omega t)+\gamma'_0}+\alpha^2e^{-k'(x+\rho
y+\omega t)-\gamma'_0}+\alpha^2e^{2k'(x+\rho y+\omega
t)+2\gamma'_0}\\
&&\quad\quad\quad
 +\alpha^6e^{-2k'(x+\rho y+\omega
t)-2\gamma'_0}+\cdots\\
&&\quad \longrightarrow 1+e^{k'(x+\rho y+\omega t)+\gamma'_0},\ \
{\rm as}\ \ \alpha\longrightarrow 0
 \end{eqnarray*}
Then the periodic solution (3.9) convergence to well-known soliton
solution
$$u=2(\ln f)_{xx}, \ \ f=1+e^{k'(x+\rho y+\omega t)+\gamma'_0}.\eqno(3.10)$$
We only need to prove that
$$\omega\longrightarrow -4k^3\pi^2+3k\rho^2,\  \ \ {\rm as }\ \ \alpha\longrightarrow
0.\eqno(3.11)$$

In fact, it is easy to see that
\begin{eqnarray*}
&&a_{11}=-32k\pi^2(\alpha^2+4\alpha^8+\cdots),\\
&&a_{12}=1+2\alpha^2+\cdots,\\
&&a_{21}=-8\pi^2k(\alpha+9\alpha^5+\cdots),\\
&&a_{22}=2(\alpha+\alpha^5+\cdots),\\
&&b_1=512k^4\pi^4(\alpha^2+16\alpha^8+\cdots)-96k^2\pi^2\rho^2(\alpha^2+4\alpha^8+\cdots),\\
&&b_2=32k^4\pi^4(\alpha+81\alpha^5+\cdots)-24k^2\pi^2\rho^2(\alpha+9\alpha^5+\cdots).
 \end{eqnarray*}
  Therefore we have
 $$b_1a_{22}-b_2a_{12}=-32k^4\pi^4\alpha+24\pi^2k^2\rho^2\alpha+o(\alpha),\ \ \
 a_{11}a_{22}-a_{12}a_{21}=8k\pi^2\alpha+o(\alpha),$$
 which lead to (3.11).

Now we consider two-periodic wave solution of the KP equation
($N=2$). Let $\hat\rho=(k_1\rho_1, k_2\rho_2),\
\hat\omega=(k_1\omega_1, k_2\omega_2).$ Substituting (3.4) into
(3.3), we have {\footnotesize\begin{eqnarray*} &&Gf\cdot
f=\sum_{m,n\in Z^2} G(D_x,D_y,D_t)e^{2\pi i<\xi,n>+\pi i<\tau
n,n>}\cdot e^{2\pi i<\xi,m>+\pi
i<\tau m,m>}\\
&&=\sum_{m,n\in Z^2}G(2\pi
i<n-m,k>,2\pi i<n-m,\hat\rho>,2\pi i<n-m,\hat\omega>)e^{2\pi i<\xi,n+m>+\pi i (<\tau m,m>+<\tau n, n>)}\\
&&\stackrel{n+m=m'}{=}\sum_{m'\in
Z^2}\sum_{n_1,n_2=-\infty}^{\infty}G(2\pi i<2n-m',k>,2\pi
i<2n-m',\hat\rho>,2\pi
i<2n-m',\hat\omega>)\\
&&\ \ \ \ \ \ \ \ \ \ \ \ \ \times \exp(\pi i(<\tau
(n-m'),n-m'>+<\tau n, n>))
\exp(2\pi i<\xi,m'>)\\
&&\equiv\sum_{m'\in Z^2}\bar{G}(m_1', m_2') e^{2\pi i <\xi,m'>}=0.
\end{eqnarray*}}
It is easy to calculate that
{\footnotesize\begin{eqnarray*}
&&\bar{G}(m_1', m_2')\\
&& =\sum_{n_1,n_2=-\infty}^{\infty}G(2\pi i<2n-m',k>,2\pi
i<2n-m',\hat\rho>,2\pi
i<2n-m',\hat\omega>)e^{\pi i(<\tau (n-m'),n-m'>+<\tau n, n>)}\\
&&{\small =\sum_{n_1,n_2=-\infty}^{\infty}G\left(2\pi
i\sum_{j=1}^{2}(2n_j'-(m_j'-2\delta_{jl}))k_j,2\pi
i\sum_{j=1}^{2}(2n_j'-(m_j'-2\delta_{jl}))k_j\rho_j,2\pi
i\sum_{j=1}^{2}(2n_j'-(m_j'-2\delta_{jl}))k_j\omega_j\right)}\\
&&\times \exp\left\{\pi
i\sum_{j,k=1}^{2}[(n_j'+\delta_{jl})\tau_{jk}(n_k'+\delta_{kl})+((m_j'-2\delta_{jl}-n_j')+\delta_{jl})
\tau_{jk}(m_k'-2\delta_{kl}-n_k')+\delta_{kl})]\right\}, \\
&&=\left\{\begin{matrix}\bar{G}(m_1'-2,m_2')e^{2\pi
i(m_1'-1)\tau_{11}+2\pi im_2'\tau_{12}},\ \ \ l=1\cr\cr
\bar{G}(m_1',m_2'-2) e^{2\pi i(m_2'-1)\tau_{22}+2\pi
im_1'\tau_{12}},\ \ \ l=2
\end{matrix}\right.
\end{eqnarray*}}
which implies that if
$$\bar{G}(0,0)=\bar{G}(0,1)=\bar{G}(1,0)=\bar{G}(1,1)=0,$$
 then $\bar{G}(m_1',m_2')=0$ and (3.2) is an exact solution
of the KP equation. Denote
 \begin{eqnarray*}
&&a_{j1}=-4\pi^2k_1\sum_{n_1,n_2=-\infty}^{\infty}
<2n-m^j,k>(2n_1-m_1^j)\delta_j(n),\\
&&a_{j2}=-4\pi^2k_2\sum_{n_1,n_2=-\infty}^{\infty}
<2n-m^j,k>(2n_2-m_2^j)\delta_j(n)\\
&&a_{j3}=24\pi^2\sum_{n_1,n_2=-\infty}^{\infty}
<2n-m^j,k>^2\delta_j(n),\\
&&a_{j4}=\sum_{n_1,n_2=-\infty}^{\infty}\delta_j(n)\\
&& b_j=\sum_{n_1,n_2=-\infty}^{\infty}(16\pi^4
<2n-m^j,k>^4-12\pi^2<2n-m^j,\hat\rho>^2)\delta_j(n),\\
&&\delta_j(n)=e^{\pi i<\tau(n-m^j),n-m^j>+\pi i<\tau n,n>},\quad j=1, 2, 3, 4\\
&&m^1=(0,0),\ \ m^2=(1,0),\ \ m^3=(0,1),\ \ m^4=(1,1)\\
&&A=(a_{kj})_{4\times 4},\ \ b=(b_1, b_2, b_3, b_4)^T,
\end{eqnarray*}
then we have $$A\left(\begin{matrix}\omega_1\cr\omega_2\cr u_0\cr
c\end{matrix}\right)=b,\eqno(3.12)$$
 from which we obtain
$$\omega_1=\frac{\Delta_1}{\Delta},\ \ \omega_2=\frac{\Delta_2}{\Delta},\ \
 u_0=\frac{\Delta_3}{\Delta},\quad c_0=\frac{\Delta_4}{\Delta}\eqno(3.13)$$
where $\Delta=|A|$ and $\Delta_j$ are produced from $\Delta$ by
replacing its $j$th column with $b$ respectively. Therefore, we
arrive at two-periodic wave solution
$$u=u_0+2(\ln f)_{xx},\eqno(3.14)$$
where $f$ and $\omega_1, \omega_2$  are given by (3.4) and (3.13)
respectively. The properties of the solution are shown in
fig.12-fig.16. Fig.12 and fig.13 shown effect of parameters $k_1,
k_2$ on period and shape of wave. Fig.14 gives effect of parameter
$\tau_{11}, \tau_{12}, \tau_{22}$ on period and shape of wave. The
fig.15 and fig.16 show that two-periodic wave can degenerate to
one-periodic wave when $k_1$ is sufficient small.

\begin{figure}[h]
$ \ \ \ \ $  {\footnotesize $ \ \ \ \ \ \ \ \ \ \ \ \ \ \ \ \ \ \
\ $}\
  {\footnotesize $(a)$}$ \ \ \ \ \ \ \ \ \ \ \ \  \ \ \ \ \ \ \ \ \ \ \ \ \
   \ \ \ \ \ \ \ \ \ \ \ \ \ \ \ \ \ \ \ \ \ \ \ \ \ \ \
 \ \ \ \ $ \\[6pt]
\begin{minipage}[u]{0.48\linewidth}
\centering \epsfig{figure=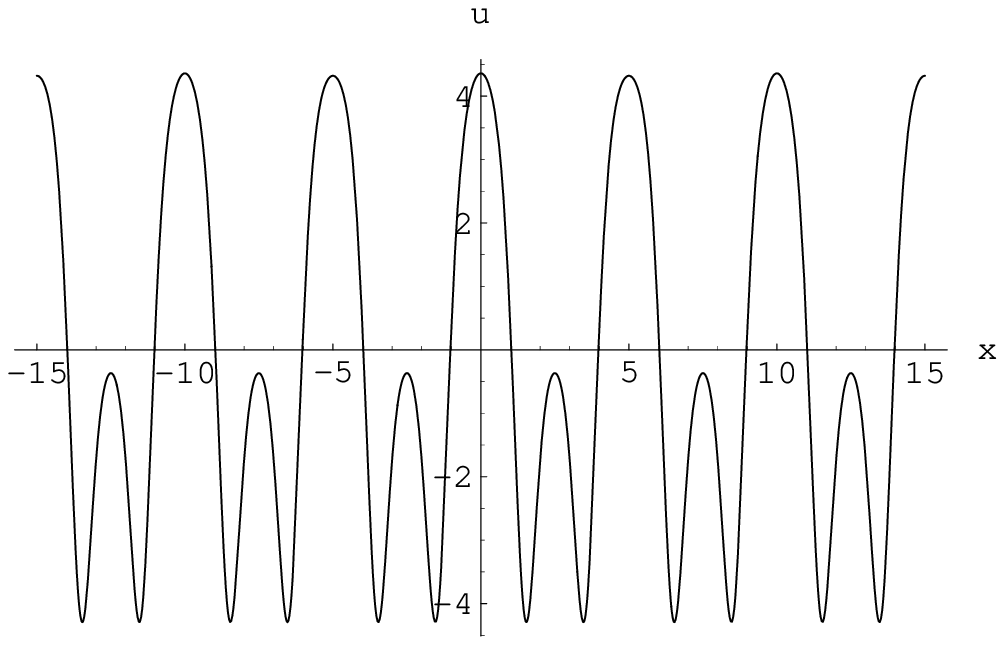,width=2.8 in }
\end{minipage}\\[12pt]
$ \ \ \ \ $  {\footnotesize $(b) \ \ \ \ \ \ \ \ \ \ \ \ \ \ \ \ \
\ \ \ \ \ \ \ \ \ \ \ \ \ \ \ \ \ \ \ \ \ \ \ \  $}\
 \ \  \ \ \ \ \ \ \
 \ \ {\footnotesize $(c)$}$ \ \ \ \ \ \ \ \ \ \ \ \  \ \ \ \ \ \ \ \ \ \ \ \ \ \ \ \ \ \ \ \ \ \ \ \ \ \ \
 \ \ \ \ $ \\[6pt]
\begin{minipage}[u]{0.48\linewidth}
\centering \epsfig{figure=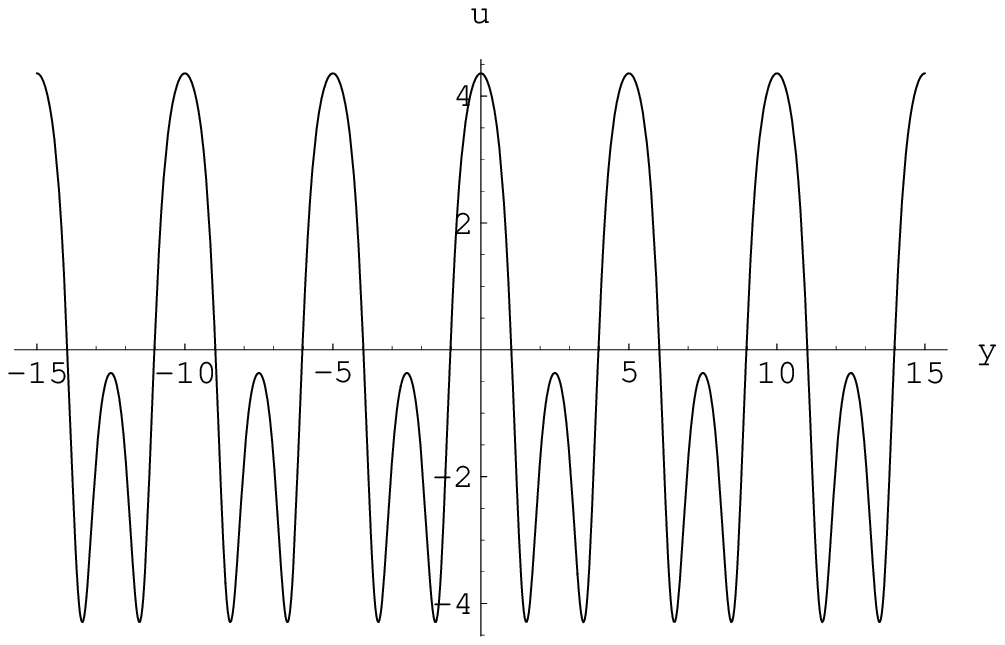,width=2.8 in }
\end{minipage}
\begin{minipage}[v]{0.48\linewidth}
\centering \epsfig{file=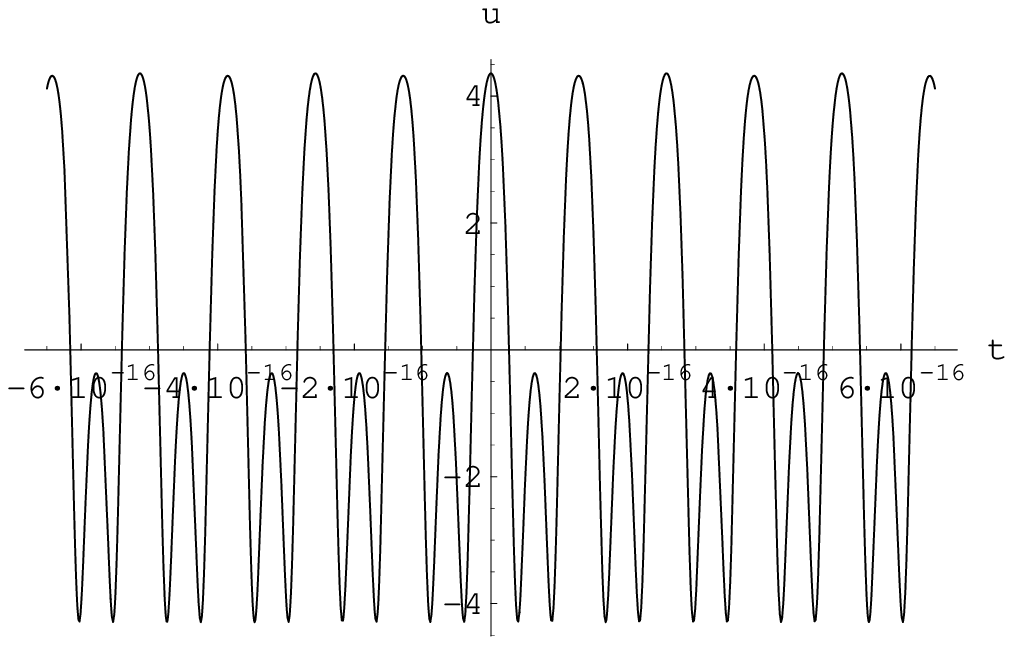,width=2.8 in}
\end{minipage}
\begin{center}
\vspace{6mm}
 {\footnotesize Figure 12. Two-periodic wave
for KP equation and the effect of parameters on wave shape: \\ (a)
along x-axis,  (b) along y-axis, (c) along t-axis, where $k_1=0.2,
k_2=-0.3,$ $\tau_{11}=0.1i, \tau_{12}=0.2i, \tau_{22}=3i,
\rho_1=1, \rho_2=2$.}
\end{center}
\end{figure}

\begin{figure}[h]
$ \ \ \ \ $  {\footnotesize $ \ \ \ \ \ \ \ \ \ \ \ \ \ \ \ \ \ \
\ $}\
  {\footnotesize $(a)$}$ \ \ \ \ \ \ \ \ \ \ \ \  \ \ \ \ \ \ \ \ \ \ \ \ \
   \ \ \ \ \ \ \ \ \ \ \ \ \ \ \ \ \ \ \ \ \ \ \ \ \ \ \
 \ \ \ \ $ \\[6pt]
\begin{minipage}[u]{0.48\linewidth}
\centering \epsfig{figure=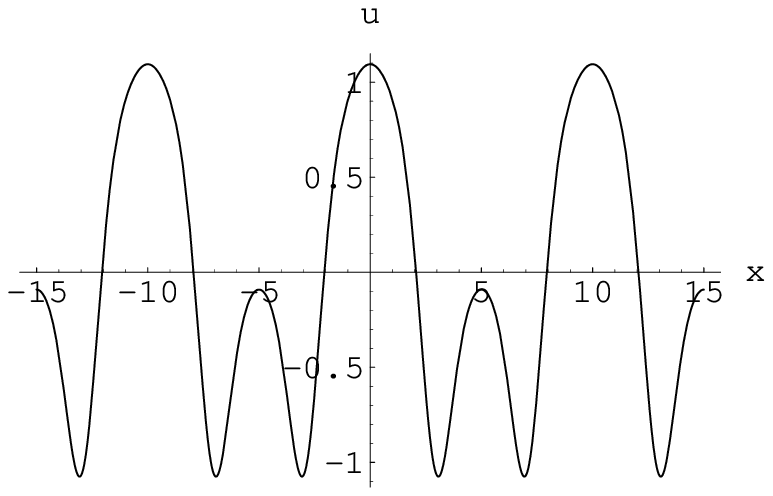,width=2.8 in }
\end{minipage}\\[12pt]
$ \ \ \ \ $  {\footnotesize $(b) \ \ \ \ \ \ \ \ \ \ \ \ \ \ \ \ \
\ \ \ \ \ \ \ \ \ \ \ \ \ \ \ \ \ \ \ \ \ \ \ \  $}\
 \ \  \ \ \ \ \ \ \
 \ \ {\footnotesize $(c)$}$ \ \ \ \ \ \ \ \ \ \ \ \  \ \ \ \ \ \ \ \ \ \ \ \ \ \ \ \ \ \ \ \ \ \ \ \ \ \ \
 \ \ \ \ $ \\[6pt]
\begin{minipage}[u]{0.48\linewidth}
\centering \epsfig{figure=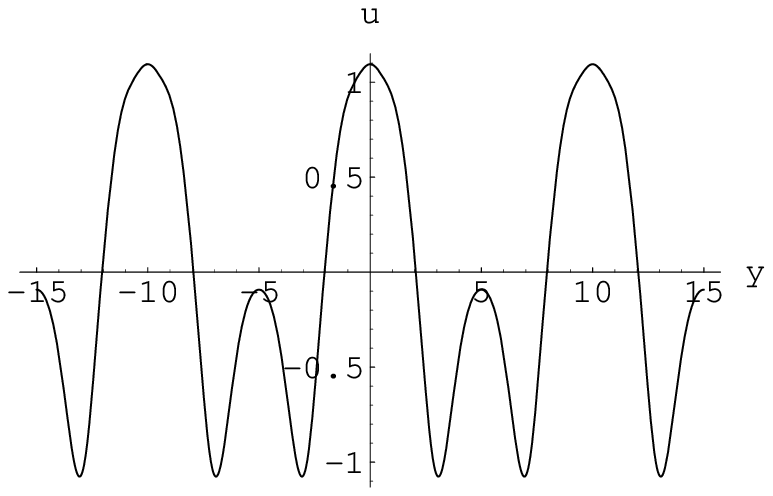,width=2.8 in }
\end{minipage}
\begin{minipage}[v]{0.48\linewidth}
\centering \epsfig{file=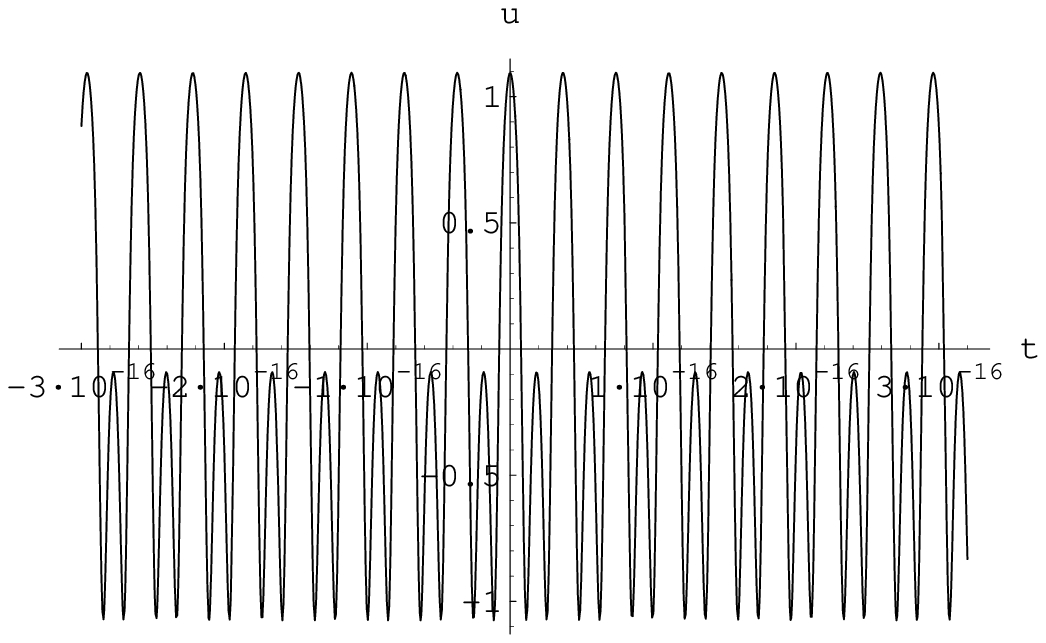,width=2.8 in}
\end{minipage}
\begin{center}
\vspace{6mm}
 {\footnotesize Figure 13. Two-periodic wave
for KP equation and the effect of parameters on wave shape: \\ (a)
along x-axis,  (b) along y-axis, (c) along t-axis, where $k_1=0.1,
k_2=-0.3,$ $\tau_{11}=0.1i, \tau_{12}=0.2i, \tau_{22}=3i,
\rho_1=1, \rho_2=2$.}
\end{center}
\end{figure}

\begin{figure}[h]
$ \ \ \ \ $  {\footnotesize $ \ \ \ \ \ \ \ \ \ \ \ \ \ \ \ \ \ \
\ $}\
  {\footnotesize $(a)$}$ \ \ \ \ \ \ \ \ \ \ \ \  \ \ \ \ \ \ \ \ \ \ \ \ \
   \ \ \ \ \ \ \ \ \ \ \ \ \ \ \ \ \ \ \ \ \ \ \ \ \ \ \
 \ \ \ \ $ \\[6pt]
\begin{minipage}[u]{0.48\linewidth}
\centering \epsfig{figure=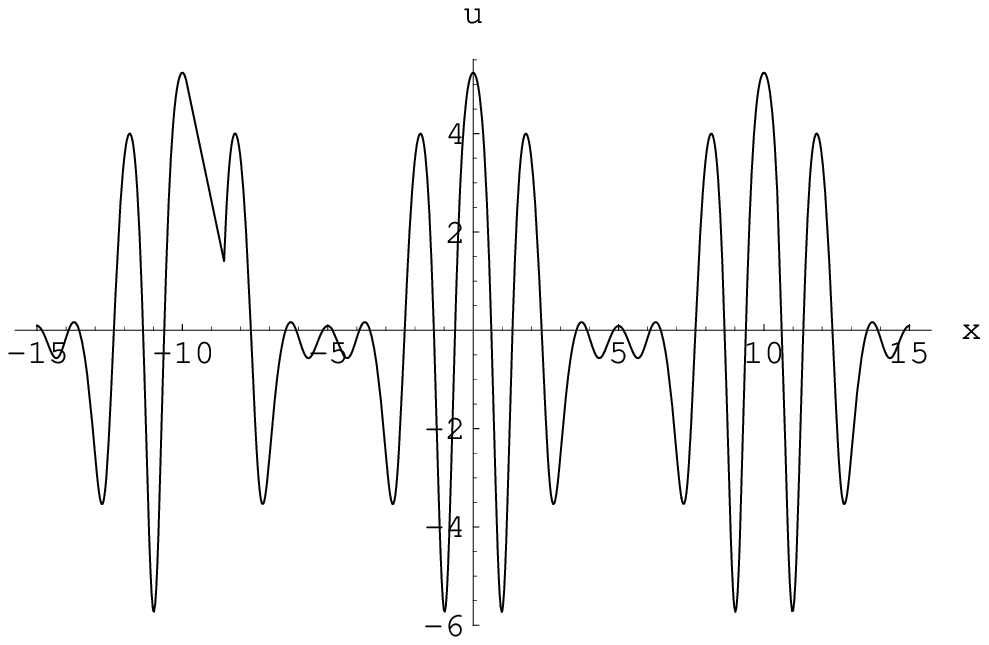,width=2.8 in }
\end{minipage}\\[12pt]
$ \ \ \ \ $  {\footnotesize $(b) \ \ \ \ \ \ \ \ \ \ \ \ \ \ \ \ \
\ \ \ \ \ \ \ \ \ \ \ \ \ \ \ \ \ \ \ \ \ \ \ \  $}\
 \ \  \ \ \ \ \ \ \
 \ \ {\footnotesize $(c)$}$ \ \ \ \ \ \ \ \ \ \ \ \  \ \ \ \ \ \ \ \ \ \ \ \ \ \ \ \ \ \ \ \ \ \ \ \ \ \ \
 \ \ \ \ $ \\[6pt]
\begin{minipage}[u]{0.48\linewidth}
\centering \epsfig{figure=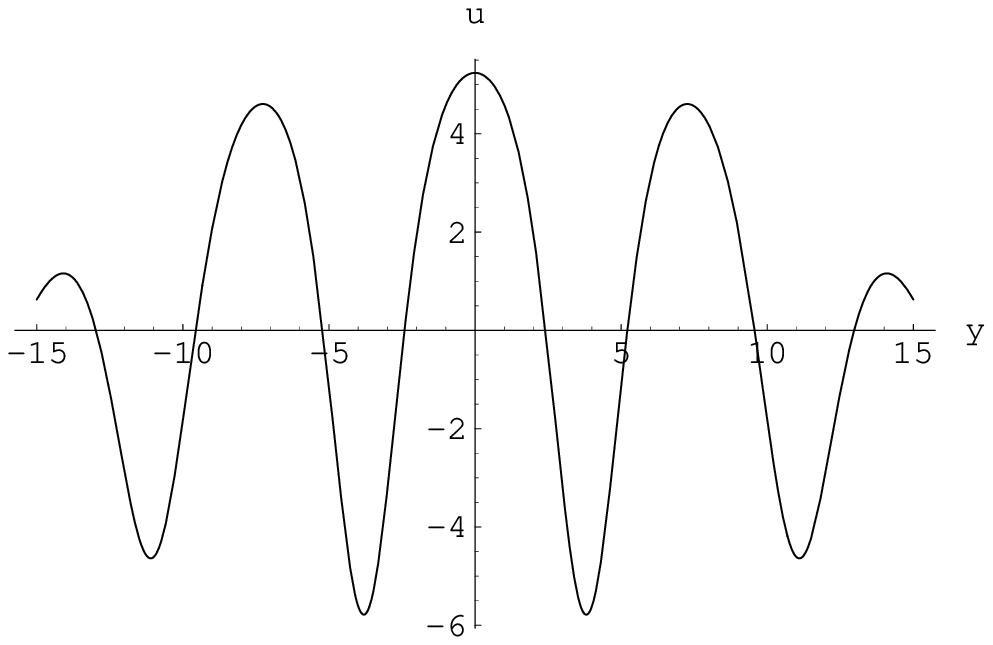,width=2.8 in }
\end{minipage}
\begin{minipage}[v]{0.48\linewidth}
\centering \epsfig{file=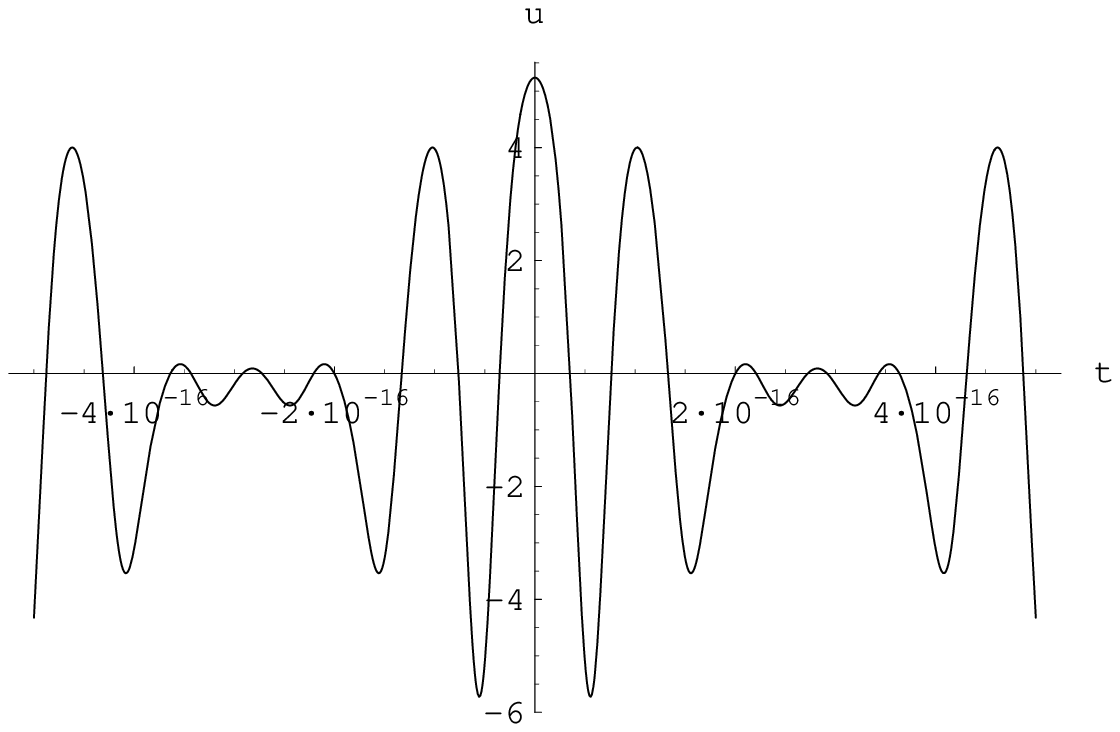,width=2.8 in}
\end{minipage}
\begin{center}
\vspace{6mm}
 {\footnotesize Figure 14. Two-periodic wave
for KP equation and the effect of parameters on wave shape: \\ (a)
along x-axis,  (b) along y-axis, (c) along t-axis, where $k_1=0.2,
k_2=-0.3, $  $\tau_{11}=0.1i, \tau_{12}=0.2i, \tau_{22}=i,
\rho_1=0.2, \rho_2=0.3$.}
\end{center}
\end{figure}

\begin{figure}[h]
$ \ \ \ \ $  {\footnotesize $ \ \ \ \ \ \ \ \ \ \ \ \ \ \ \ \ \ \
\ $}\
  {\footnotesize $(a)$}$ \ \ \ \ \ \ \ \ \ \ \ \  \ \ \ \ \ \ \ \ \ \ \ \ \
   \ \ \ \ \ \ \ \ \ \ \ \ \ \ \ \ \ \ \ \ \ \ \ \ \ \ \
 \ \ \ \ $ \\[6pt]
\begin{minipage}[u]{0.48\linewidth}
\centering \epsfig{figure=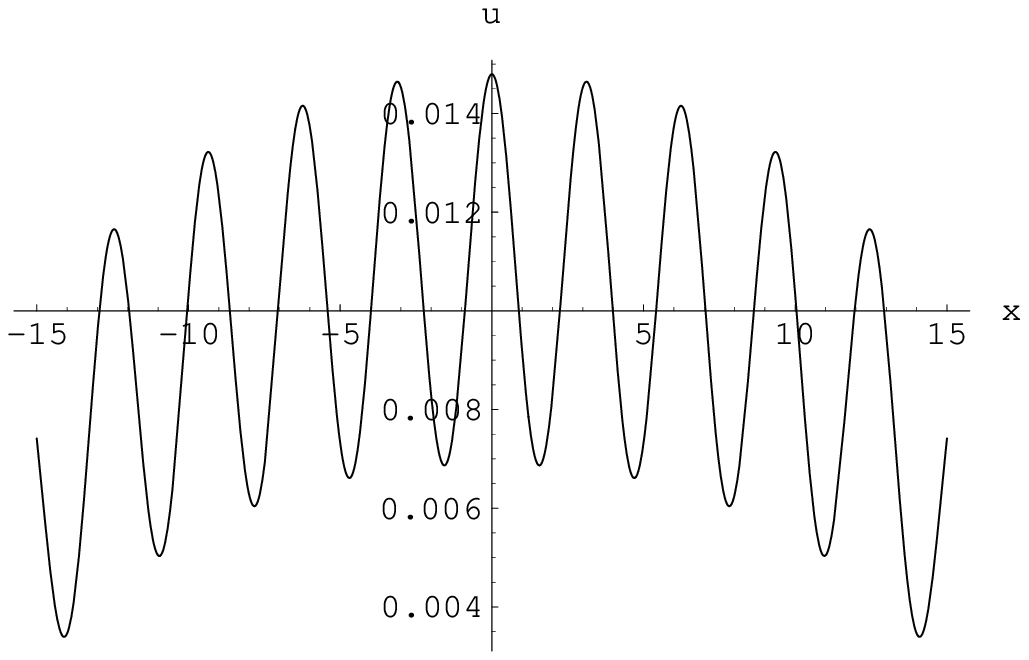,width=2.8 in }
\end{minipage}\\[12pt]
$ \ \ \ \ $  {\footnotesize $(b) \ \ \ \ \ \ \ \ \ \ \ \ \ \ \ \ \
\ \ \ \ \ \ \ \ \ \ \ \ \ \ \ \ \ \ \ \ \ \ \ \  $}\
 \ \  \ \ \ \ \ \ \
 \ \ {\footnotesize $(c)$}$ \ \ \ \ \ \ \ \ \ \ \ \  \ \ \ \ \ \ \ \ \ \ \ \ \ \ \ \ \ \ \ \ \ \ \ \ \ \ \
 \ \ \ \ $ \\[6pt]
\begin{minipage}[u]{0.48\linewidth}
\centering \epsfig{figure=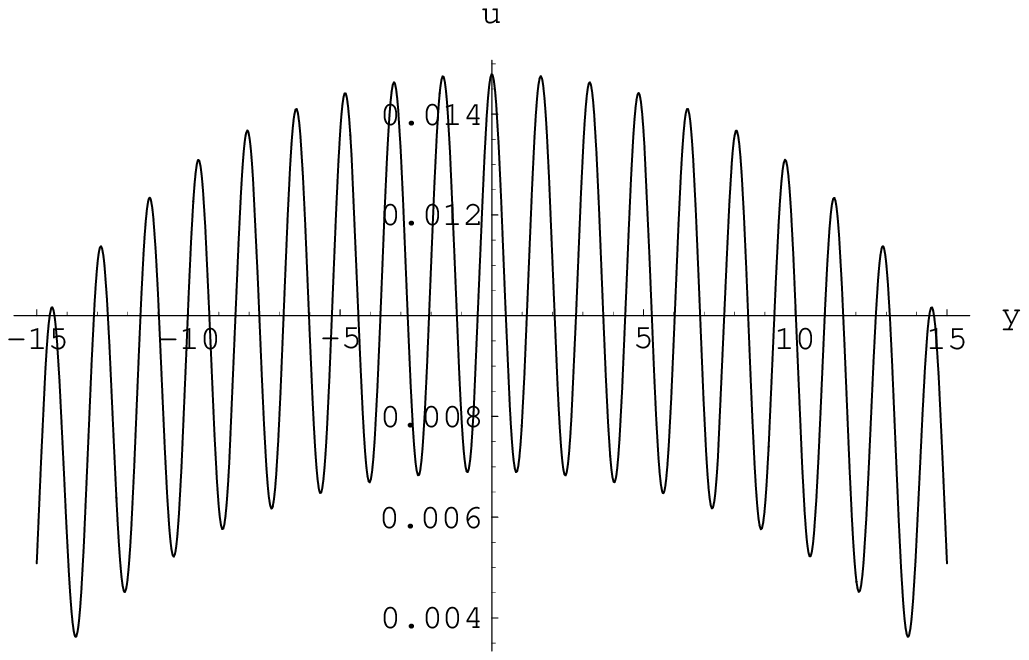,width=2.8 in }
\end{minipage}
\begin{minipage}[v]{0.48\linewidth}
\centering \epsfig{file=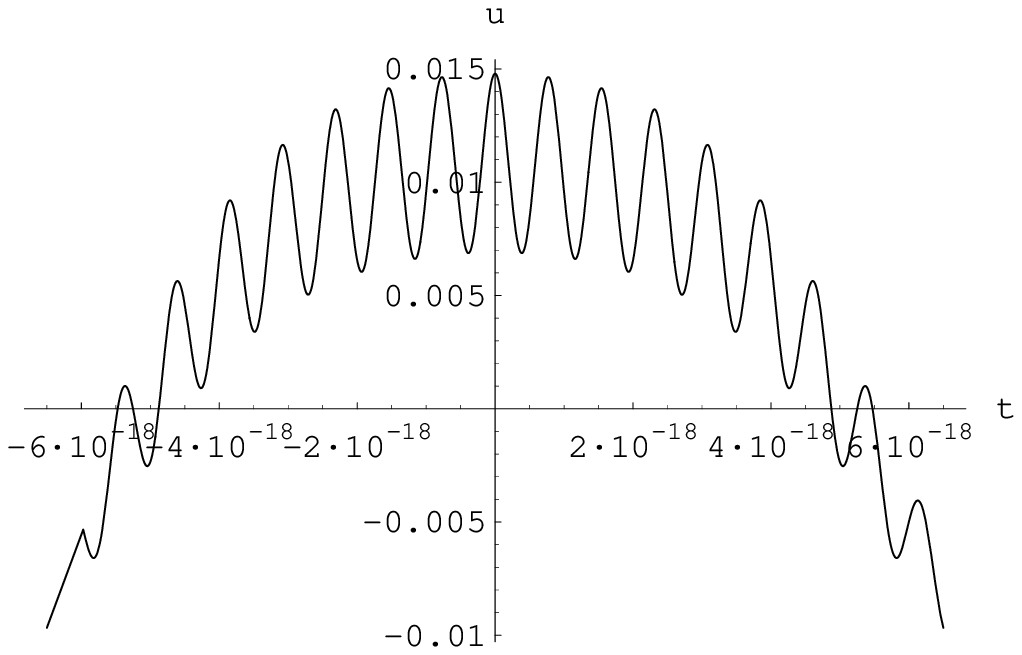,width=2.8 in}
\end{minipage}
\begin{center}
\vspace{6mm}
 {\footnotesize Figure 15. Two-periodic wave
for KP equation and the effect of parameters on wave shape: \\ (a)
along x-axis,  (b) along y-axis, (c) along t-axis, where
$k_1=0.01, k_2=-0.3, $  $\tau_{11}=0.1i, \tau_{12}=0.2i,
\tau_{22}=i, \rho_1=0.2, \rho_2=0.3$.}
\end{center}
\end{figure}

\begin{figure}[h]
$ \ \ \ \ $  {\footnotesize $ \ \ \ \ \ \ \ \ \ \ \ \ \ \ \ \ \ \
\ $}\
  {\footnotesize $(a)$}$ \ \ \ \ \ \ \ \ \ \ \ \  \ \ \ \ \ \ \ \ \ \ \ \ \
   \ \ \ \ \ \ \ \ \ \ \ \ \ \ \ \ \ \ \ \ \ \ \ \ \ \ \
 \ \ \ \ $ \\[6pt]
\begin{minipage}[u]{0.48\linewidth}
\centering \epsfig{figure=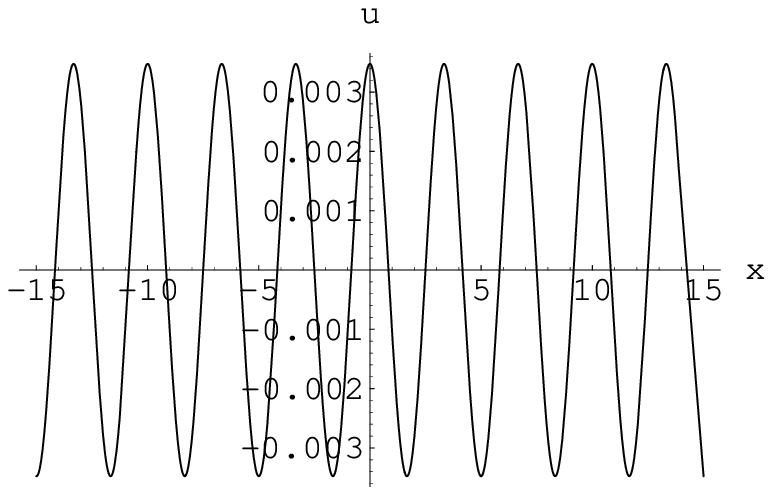,width=2.8 in }
\end{minipage}\\[12pt]
$ \ \ \ \ $  {\footnotesize $(b) \ \ \ \ \ \ \ \ \ \ \ \ \ \ \ \ \
\ \ \ \ \ \ \ \ \ \ \ \ \ \ \ \ \ \ \ \ \ \ \ \  $}\
 \ \  \ \ \ \ \ \ \
 \ \ {\footnotesize $(c)$}$ \ \ \ \ \ \ \ \ \ \ \ \  \ \ \ \ \ \ \ \ \ \ \ \ \ \ \ \ \ \ \ \ \ \ \ \ \ \ \
 \ \ \ \ $ \\[6pt]
\begin{minipage}[u]{0.48\linewidth}
\centering \epsfig{figure=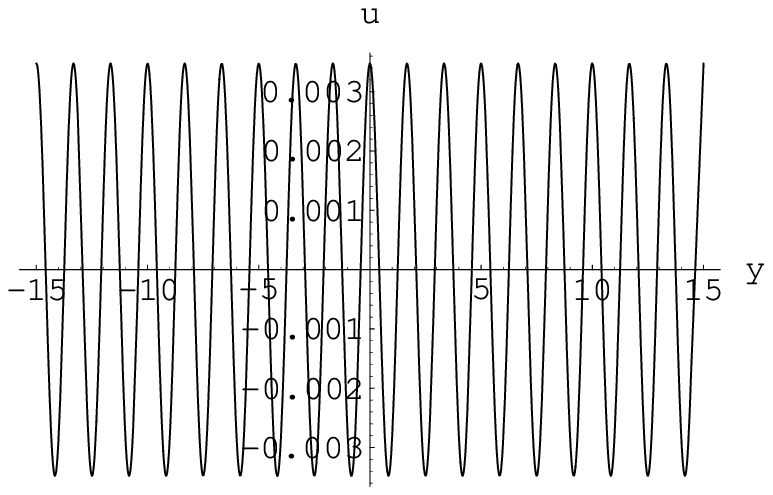,width=2.8 in }
\end{minipage}
\begin{minipage}[v]{0.48\linewidth}
\centering \epsfig{file=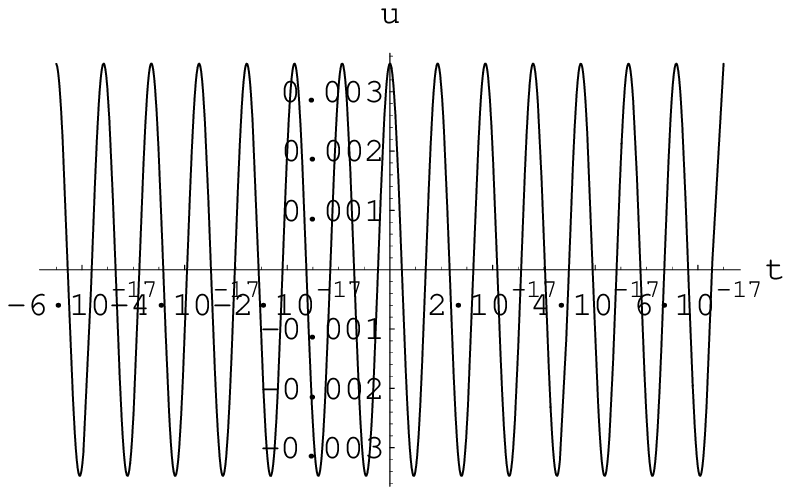,width=2.8 in}
\end{minipage}
\begin{center}
\vspace{6mm}
 {\footnotesize Figure 16. Two-periodic wave
for KP equation and the effect of parameters on wave shape: \\ (a)
along x-axis,  (b) along y-axis, (c) along t-axis, where
$k_1=0.001, k_2=-0.3, $  $\tau_{11}=0.1i, \tau_{12}=0.2i,
\tau_{22}=i, \rho_1=0.2, \rho_2=0.3$.}
\end{center}
\end{figure}

Two-soliton solution of the KP equation can be obtained as limit
of the periodic solution (3.14).  We write $f$ as
\begin{eqnarray*}
&& f=1+(e^{2\pi i\xi_1}+e^{-2\pi i\xi_1})e^{\pi i\tau_{11}}
+(e^{2\pi i\xi_2}+e^{-2\pi i\xi_2})e^{\pi
i\tau_{22}}\\
&&+(e^{2\pi i(\xi_1+\xi_2)}+e^{-2\pi i(\xi_1+\xi_2)})e^{\pi
i(\tau_{11}+2\tau_{12}+\tau_{22})}+\cdots
\end{eqnarray*}
Setting $\gamma'_1=\gamma_1+\frac{1}{2}\tau_{11},\
\gamma'_2=\gamma_2+\frac{1}{2}\tau_{22},\ \xi_1'=2\pi i\xi_1-\pi
i\tau_{11},\ \ \xi_2'=2\pi i\xi_2-\pi i\tau_{22},
 \ \ \tau_{12}=i{\tau}'$ (${\tau}'$ is a real),
 we get
\begin{eqnarray*}
&&f=1+e^{\xi_1'}+e^{\xi_2'}+e^{\xi_1'+\xi_2'+2\pi i\tau_{12}
}+\alpha_1^2e^{-\xi_1'}
 +\alpha_2^2e^{-\xi_2'}+\alpha_1^2\alpha_2^2e^{-\xi_1-\xi_2+2\pi i\tau_{11}}+\cdots\\
&&\longrightarrow 1+e^{\xi_1'}+e^{\xi_2'}+e^{\xi_1'+\xi_2'-2\pi
{\tau'}},\ \ {\rm as}\ \ \alpha_1, \alpha_2
\longrightarrow 0
 \end{eqnarray*}
where
$$\alpha_1=e^{\pi i\tau_{11}},\ \ \alpha_2=e^{\pi
i\tau_{22}}, \ \ \xi_j'=2\pi
i[k_j(x+\rho_j+\omega_jt)+\gamma'_j],$$
$$ e^{-2\pi
{\tau'}}=\frac{(k_1-k_2)^2-(\rho_1-\rho_2)^2}{(k_1+k_2)^2-(\rho_1-\rho_2)^2},\
\ \  \omega_i'\rightarrow  -4k_j^3\pi^2+3k\rho_j^2,\ \ \ \ {\rm
as}\ \ \ \ \alpha_1, \alpha_2\rightarrow0.$$
\\[12pt]
{\leftline{\normalsize {\bf Acknowledgments}}}

The work described in this paper was partially supported by a
grant from City University of Hong Kong (Project No. 9040967).


\newpage

\begin{thebibliography}{99}
\footnotesize
 \baselineskip=18pt
 \bibitem{ab} M.J. Ablowitz and  P.A. Clarkson, Solitons, Nonlinear Evolution
           Equations and Inverse Scattering, Cambridge University Press, 1991.

\bibitem{be}  R. Beals and R.R. Coifman, Comm. pure. Appl. Math. 37 (1984) 39.

\bibitem{}    V.B. Matveev and M.A. Salle, Darboux transformation and solitons,
            Springer,  Berlin, 1991.

\bibitem{}  C.H. Gu, H.S. Hu and Z.X. Zhou, Darboux Transformations in Soliton
             Theory and its Geometric Applications,Shanghai Science Technology
              Publisher, Shanghai, 1999.

\bibitem{}  S.B. Leble and N.V. Ustinov, J. Phys. A  26 (1993) 5007.

\bibitem{}  P.G. Esteevez, J. Math. Phys. 40 (1999) 1406.

\bibitem{} R. Hirota and J. Satsuma, Prog. Theor. Phys. {\bf 57} (1977) 797.
\bibitem{} R. Hirota, Direct methods in soliton theory, Springer, 2004.
\bibitem{} X. B. Hu, P. A. Clarkson, J. Phys. A,{\bf 28}(1995), 5009.
\bibitem{} X. B. Hu, C X Li, J.J. C. Nimmo, G. F. Yu, J. Phys A,
{\bf 38}(2005), 195.
\bibitem{} R Hirota, Y. Ohta,  J. Phys Soc. Jpn {\bf 60}(1991) 798
\bibitem{} J X Zhao, C X Li, X. B Hu,  J. Phys Soc. Jpn {\bf 73}(2004) 1159

\bibitem{}  B. A. Dubrovin, Funct. Anal. Appl. {\bf 9} (1975) 41.
\bibitem{}  A. Its and V. Matveev, Funct. Anal. Appl. {\bf 9} (1975) 69.
\bibitem{}  E. Belokolos, A. Bobenko, V. Enol'skij, A, Its and V. Matveev,
            {\it Algebro-Geometrical Approach to Nonlinear Integrable
              Equations}, Springer, Berlin, 1994.
\bibitem{}  S.P. Novikov, and S.V. Manakov, L.P. Pitaeskii and V.E. Zakharov,
    {\it Theory of Solitons, The Snverse Scattering Methods}, Nauka, Moscow, 1980.
\bibitem{}  P. L. Christiansen, J. C. Eilbeck, V. Z. Enolskii and N. A.
            Kostov,  Proc. R. Soc. London A Math.  {\bf 451} (1995) 685
\bibitem{}  R. G. Zhou, J. Math. Phys. {\bf 38} (1997) 2535.
\bibitem{}  C. W. Cao, Y. T. Wu and X. G. Geng, J. Math. Phys. {\bf 40} 1999, 3948.
\bibitem{}  X. G. Geng and Y. T. Wu, J. Math. Phys. {\bf 40} (1999), 2971.
\bibitem{}  X. G. Geng,  C. W. Cao and H. H. Dai, J. Phys. A, {\bf 34} 2001, 989.
\bibitem{} Z. J. Qiao, Reviews in Math. Phys. {\bf 13} (2001), 545.
\bibitem{}  C. W. Cao,  X. G. Geng and H. Y. Wang, J. Math. Phys. {\bf 43} 2002, 621.
\bibitem{}   H. H. Dai and X. G. Geng, Chaos, Solitons and Fractrals, {\bf 18}(2003), 1031
\bibitem{}   B. Deconinck, M. V. Hoeij, Physica D, {\bf 152-153}(2001)  28
\bibitem{}    B Docninck, M Meil, A. Bobenko, M. V. Hoeij and M. Schmies,
             Math. of Comput. 1417.
\end{thebibliography}
\end{document}